\documentclass[lettersize,journal]{IEEEtran}
\usepackage{amsmath,amsfonts}
\usepackage{booktabs}
\usepackage{algorithmic}
\usepackage{algorithm}
\usepackage{array}
\usepackage[caption=false,font=normalsize,textfont=sf]{subfig}
\usepackage{textcomp}
\usepackage{stfloats}
\usepackage{url}
\usepackage{verbatim}
\usepackage{graphicx}
\usepackage{cite}
\usepackage{multicol}
\usepackage{amssymb}
\usepackage{lineno}
\usepackage{makecell}
\usepackage{indentfirst}
\usepackage{bm}
\usepackage{color}
\usepackage{cite}
\usepackage{epstopdf}
\usepackage{amssymb}
\usepackage{xcolor}
\usepackage{tikz}
\usepackage{mathtools}
\usepackage{pifont}
\newcommand{\tr}{\mathrm{tr}}
\newcommand{\HH}{\mathrm{H}}	
\newcommand{\TT}{\mathrm{T}}

\newtheorem{remark}{Remark}
\newtheorem{corollary}{Corollary}
\newtheorem{proposition}{Proposition}

\newtheorem{lemma}{Lemma}
\begin{document}
	
	\title{Sensing Mutual Information with Random Signals  in Gaussian Channels: Bridging Sensing and Communication Metrics}
	
	%A Deterministic Equivalent for 
	\author{%Lei Xie, Fan Liu, Mingxuan Zheng, Zhanyuan Xie, Zheng Jiang, and Shenghui Song
		Lei Xie, Fan Liu, Jiajin Luo, and Shenghui Song
		%\thanks{L. Xie is with the School of Cyber Science and Engineering, Southeast University, China. e-mail: ()}
		\thanks{Part of this paper will be presented at  IEEE International Conference on Communications 2024 \cite{xie2023sensing}.}
		\thanks{L. Xie and S. Song are with Department of Electronic and Computer Engineering, the Hong Kong University of Science and Technology, Hong Kong. %e-mail: ($\{$eelxie, eeshsong$\}$@ust.hk). 
			F. Liu is with Southern University of Science and Technology, China. 
			J. Luo is with Huawei Technologies Company Ltd.}
	}
	
	%\author{Lei~Xie$^{\star}$, Fan Liu$^{\dagger}$, Mingxuan Zheng, Zhanyuan Xie$^{\ddagger}$, Zheng Jiang$^{\ddagger}$, and Shenghui Song$^{\star}$\\
		%{$^{\star}$The Hong Kong University of Science and Technology, Hong Kong}\\
		%{$^{\dagger}$Southern University of Science and Technology, China}\\
		%{$^{\ddagger}$China Telecom Research Institute, China}
		%%{Email: $\{$eelxie, eeshsong$\}$@ust.hk}
		%}

	\maketitle
	
	\begin{abstract}
		Sensing performance is typically evaluated by classical radar metrics, such as Cramér-Rao bound and signal-to-clutter-plus-noise ratio. 
		The recent development of the integrated sensing and communication (ISAC) framework motivated the efforts to unify the performance metric for sensing and communication, where mutual information (MI) was proposed as a sensing performance metric with \emph{deterministic} signals. 
		%and researchers have proposed to utilize mutual information (MI) to measure the sensing performance with \emph{deterministic} signals. 
		However, the need of communication in ISAC systems necessitates the transmission of \emph{random} signals for sensing applications, whereas an explicit evaluation for the sensing mutual information (SMI) with random signals is not yet available in the literature. 
		This paper aims to fill the research gap and investigate the unification of sensing and communication performance metrics. 
		%This paper investigates the achievable performance and precoding design for ISAC systems employing random signals. 
		For that purpose, we first derive the explicit expression for the SMI with random signals utilizing random matrix theory. 
		On top of that, we further build up the connections between SMI and traditional sensing metrics, such as ergodic minimum mean square error (EMMSE), ergodic linear minimum mean square error (ELMMSE), and ergodic Bayesian Cram\'{e}r-Rao bound (EBCRB). 
		Such connections open up the opportunity to unify sensing and communication performance metrics, which facilitates the analysis and design for ISAC systems. 
		Finally, SMI is utilized to optimize the precoder for both sensing-only and ISAC applications. 
		Simulation results validate the accuracy of the theoretical results and the effectiveness of the proposed precoding designs. 
	\end{abstract}
	
	\begin{IEEEkeywords}
		Integrated sensing and communication, sensing mutual information, random signals, sensing degree of freedom, precoding design.
	\end{IEEEkeywords}
	
	\section{Introduction}
	
	With the development of innovative applications that demand accurate environmental information, e.g., autonomous driving and unmanned aerial vehicle (UAV) networks, sensing becomes an essential requirement for future wireless networks. To this end, the integrated sensing and communications (ISAC) framework has attracted tremendous attention  \cite{liu2020joint,liu2022integrated,liu2022survey,xie2023collaborative,xie2023networked}.  
	However, there are still substantial challenges that need to be tackled before we can fully unleash the potential of ISAC. One of the most significant issues is the distinct performance evaluation metrics for sensing and communication. 	
	%. For example, sensing and communication systems have been developed separately for many years with their own performance metrics. 
	In particular, sensing performance is typically characterized by metrics such as Cram\'{e}r-Rao bound (CRB) \cite{liu2021cramer}, signal-to-clutter-plus-noise ratio \cite{xie2022perceptive}, and minimum mean-square error (MMSE) \cite{herbert2017mmse}, while communication performance is usually evaluated by the achievable rate \cite{hachem2008new}, bit error rate \cite{jeruchim1984techniques}, and outage probability \cite{ko2000outage}. 
	%usually measured by mutual information (MI) \cite{hachem2008new} and outage probability \cite{ko2000outage}. 
	The inconsistency in performance metrics between two systems poses challenges to the design of ISAC systems.
	
	Recently, some research efforts have been devoted to unifying the performance metrics for sensing and communication.
	For that purpose, mutual information (MI), a metric commonly used to measure communication performance, has also been utilized to evaluate sensing performance \cite{bell1993information,tang2018spectrally,yang2007mimo}. 
	%performance of both sensing and communication systems. %since it can quantify the amount of information that can be transmitted from source to receiver. 
	%In fact, MI has been widely utilized to measure the maximum information exchange between the transceivers. Meanwhile, MI can be also utilized to measure sensing performance but its physical meaning is not clear  \cite{bell1993information,tang2018spectrally,yang2007mimo}. 
	The authors of \cite{bell1993information} demonstrated that the waveform that maximizes the MI between the random target response and the received signal can also lead to enhanced sensing performance. The authors of \cite{yang2007mimo} showed that the radar waveform that maximizes the MI between the Gaussian-distributed target response and the received signal also minimizes the MMSE for estimating the target response. 
	
	However, existing works  \cite{bell1993information,tang2018spectrally,yang2007mimo}  mainly focused on sensing mutual information (SMI) with deterministic signals, which have been widely employed in sensing applications, due to some favorable properties: 
	1) deterministic signals allow for precise control of key parameters, such as amplitude and phase of waveform, which enables accurate parameter estimation. For example, the radar using frequency modulated continuous wave (FMCW) can provide accurate range and velocity information \cite{li200824,wang2014application}; and
	2) deterministic signals facilitate some important techniques in radar systems, such as pulse compression and coherent integration, which enhance range resolution and target detection ability \cite{huang2019long,xie2020recursive}.
	%3) Deterministic signals enable accurate target identification and discrimination. By using specific waveform schemes, radar systems can distinguish different targets \cite{743818}.
	
	\begin{figure}[!t]
		\centering
		\includegraphics[width=3.5in]{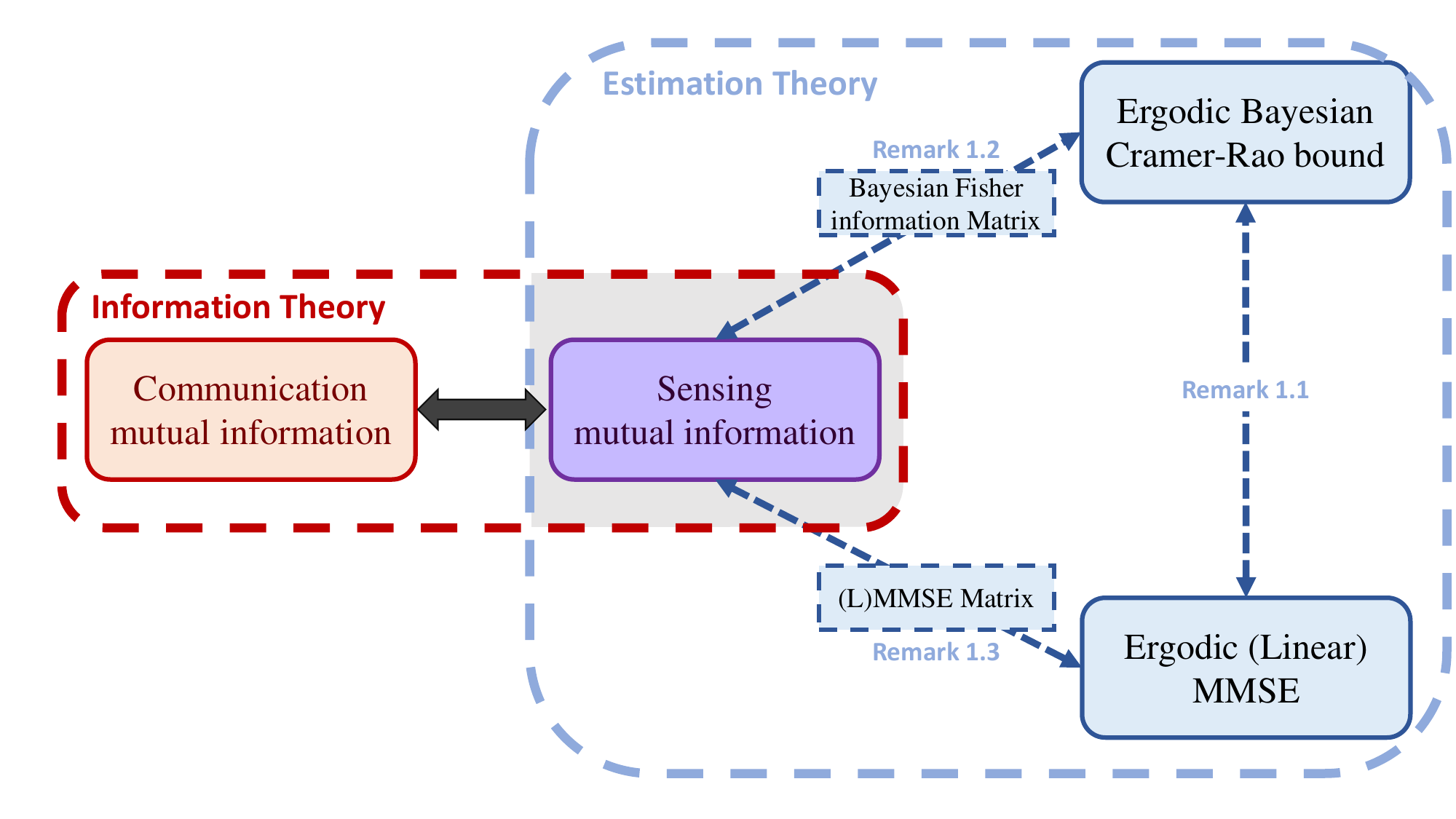}
		\caption{Illustration of the connection between SMI and other communication/sensing metrics.}
		\label{illu_connection}
	\end{figure}
	
	In contrast to radar systems, communication applications require the signals to be random. As a result, it is crucial to evaluate the sensing performance with random signals in ISAC systems  \cite{liu2023deterministic,dong2023rethinking,xiong2023fundamental,xiong2023generalized,lu2023sensing,lu2023random}. However, the explicit evaluation for SMI with random signals remains widely unexplored, and in fact, even the evaluation for traditional sensing metrics with random signals is also missing. 	
	%As a result, there is an intrinsic conflict between sensing and communication in ISAC systems, which poses significant challenges on the sensing performance evaluation in ISAC systems to handle the randomness of transmitted ISAC signals \cite{liu2023deterministic,dong2023rethinking,xiong2023fundamental,xiong2023generalized,lu2023sensing,lu2023random,xie2023sensing}.
	%The competition between sensing and communication lead to the deterministic-random tradroff (DRT) in ISAC systems 
	%This tradeoff arises due to the limited resources and shared bandwidth in ISAC systems, where both sensing and communication tasks need to be performed using the same signals. 
	%Some works \cite{liu2023deterministic,dong2023rethinking,xiong2023fundamental,xiong2023generalized} investigated the DRT from a rate-distortion perspective, which suggested that advanced precoding design techniques played a significant role in optimizing the DRT.
	%However, the randomness of transmitted signals poses significant challenges on the sensing performance evaluation in ISAC systems. 
	To this end, some researchers resorted to approximations or numerical methods \cite{xiong2023fundamental,xiong2023generalized,lu2023sensing,lu2023random}. 
	For example, the authors of \cite{xiong2023fundamental} derived both lower and upper bounds for the Bayesian CRB with random signals, namely, the ergodic Bayesian CRB (EBCRB).  
	In \cite{lu2023sensing,lu2023random}, the ergodic linear MMSE (ELMMSE), which characterizes the estimation error of the target response matrix averaged over random signals, was studied for sensing system.  %Based on this, a data-dependent scheme for precoding design was proposed to minimize the ELMMSE. To reduce the computational cost, a data-independent precoding was developed by exploiting the stochastic gradient projection. 
	%Then, a data-dependent scheme for precoding design was proposed to minimize the ELMMSE by exploiting the stochastic gradient projection (SGP). However, the explicit expressions for the sensing metrics with random signals are often challenging to derive and are not yet available in the literature, though they are crucial for analyzing and optimizing sensing performance. 

	This paper aims to unify the performance metrics for sensing and communication in ISAC systems. For that purpose, we first derive an explicit expression for the SMI between the random target response and the received signals with random transmitting signals. 
	It is shown that SMI can be interpreted as the maximum information that can be exacted from the target response matrix by the sensing receiver. 
	Next, by studying the relation between SMI and other traditional sensing metrics, e.g., ergodic MMSE (EMMSE), ELMMSE, and EBCRB, we demonstrate that SMI can serve as a bridge to connect communication and sensing metrics, as illustrated in Fig. \ref{illu_connection}.  
	One interesting finding is that SMI is directly related to Bayesian Fisher information and maximizing SMI is equivalent to minimizing the EBCRB, which is the lower bound for the estimation error of Bayesian estimators. Moreover, the estimation error of MMSE and LMMSE estimators approaches the BCRB in the case that the target response matrix is Gaussian-distributed. %These connections highlight the importance of SMI and provides valuable insights into the design of ISAC systems. 

	The connections between SMI and traditional sensing metrics, %in the context of optimization, 
	along with the widespread use of MI in measuring the performance of communication systems, 
	%the fact that MI has been widely utilized to measure the performance of communication systems, 
	motivate us to adopt MI for evaluating the communication and sensing performance in a unified manner. %Such a unified framework will facilitate analysis and design of in ISAC systems. 
	To demonstrate the benefits of such a unified framework, we investigate the precoding designs for both sensing-only and ISAC applications, by imposing MI as the performance metric. 
	Simulations results show that the proposed methods achieve superior performance than existing methods.

	The main contributions of this paper are summarized as:
	
	\begin{enumerate}
		\item %We derive the explicit expression for the SMI by exploiting the principles of random matrix theory. 
		We derive explicit expressions for SMI, EMMSE, ELMMSE, and EBCRB with random signals by leveraging random matrix theory. It is shown that the SMI with random signals is upper-bounded by that with a properly-designed deterministic signal. Furthermore, as the number of samples increases, the SMI loss caused by the utilization of random signals will diminish. 
		%Moreover, these expressions contribute to a better understanding of the deterministic and random tradeoff (DRT) between sensing and communications.  %This insight contributes to a better understanding of the trade-offs between randomness and determinism in signal design.
		%These expressions provide a clear understanding of the deterministic and random tradeoff (DRT) between sensing and communications. 
		\item We demonstrate that SMI can serve as a bridge to connect communication and sensing metrics. In particular, we build up the connections between SMI, EMMSE, ELMMSE, and EBCRB. %Furthermore, we offer a physical interpretation for SMI as the maximum amount of information about the target response matrix that can be extracted at the sensing receiver. 
		These results provide a unified framework for evaluating communication and sensing performance, facilitating the analysis and design of ISAC systems. 
		\item Building upon the theoretical results, we propose two precoding design methods for sensing-only and ISAC scenarios. Remarkably, although a fixed-point equation is involved in the computation of SMI, the explicit expression for its gradient is available, which enables efficient optimization. %The simulations demonstrate that the precoder maximizing the SMI surpasses the one maximizing its upper bound. %, which is equivalent to the SMI with infinite samples. 
		Simulation results validate the accuracy of the theoretical analysis results and effectiveness of the proposed optimization methods. 
		%These results highlight the superiority of our proposed precoding methods in achieving better performance.% and better tradeoffs between sensing and communications. 
	\end{enumerate}
	
	The remainder of the paper is organized as follows.  
	Section II presents the system model. Section III provides an explicit expression for the SMI and reveals some physical insights. 
	Section IV studies the relation between SMI and other sensing metrics and provides a physical interpretation for SMI. 
	Section V introduces two approaches to maximize the SMI by optimizing the transmitted precoder for both sensing-only and ISAC systems. 
	Simulation results are given in Section VI to validate the accuracy of the theoretical results and the effectiveness of the proposed optimization-based approaches.	
	Finally, Section VII concludes the paper, summarizing the key findings and contributions.

	%\newpage
	
	\section{System Model}
\begin{figure}[!t]
		\centering
		\includegraphics[width=3.5in]{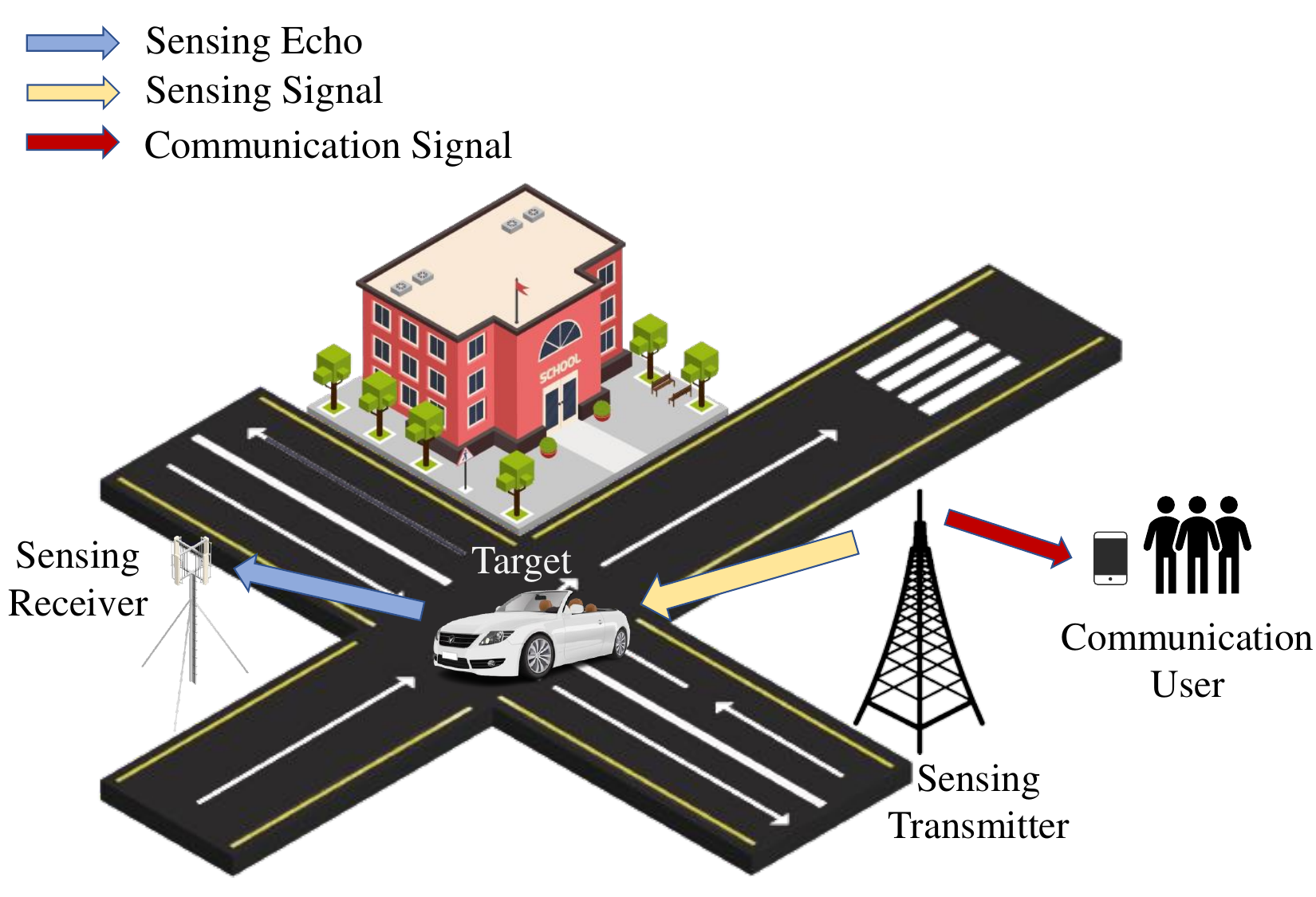}
		\caption{Illustration of the considered ISAC system.}
		\label{ill_ISAC}
	\end{figure}

	Consider an ISAC system %operating in the downlink communication and bi-static sensing manner as illustrated in Fig. \ref{ill_ISAC}, which is 
	composed of an ISAC transmitter with $N_T$ antennas, a communication receiver with $N_C$ antennas, and a sensing receiver with $N_R$ antennas as shown in Fig. \ref{ill_ISAC}. In this paper, we consider the simultaneous downlink transmission and target sensing with the bi-static sensing scheme. In the following, we will present the communication and sensing model, respectively.
	
	\subsection{Communication model}
	The received communication signal is given by
	\begin{equation}
		\begin{split}
			\mathbf{Y}_c =  \mathbf{H}_c \mathbf{X}+ \mathbf{N}_c \in \mathbb{C}^{N_C\times N_S},
		\end{split}
	\end{equation}
	where 
	%\begin{equation}
	%	\mathbf{H}_c = \sum_{k=1}^{K_c} \varepsilon_{c,k} \mathbf{b}_c(\vartheta_{c,k}) \mathbf{a}(\phi_{c,k})^\HH \in \mathbb{C}^{N_C\times N_T}.
	%\end{equation}
	$\mathbf{X}\in \mathbb{C}^{N_T\times N_S}$ represents the transmitted signal, $\mathbf{H}_c \in \mathbb{C}^{N_C\times N_T}$ denotes the MIMO communication channel, and %$\phi_{c,k}$ and $\varepsilon_{k}$ denote the AOD and pathloss corresponding to the $k$th path, respectively. Moreover,  with $\mathbb{E}(||\mathbf{H}_c||_F^2)=N_C N_T$
	$\mathbf{N}_c$ represents the additive white Gaussian noise. In this paper, we consider the widely adopted Gaussian signal with $\mathbf{X} = \mathbf{F} \mathbf{S}$, where $\mathbf{F}\in \mathbb{C}^{N_T\times N_T}$ represents the precoding matrix and $\mathbf{S}\in \mathbb{C}^{N_T\times N_S}$ denotes a random matrix whose entries are independent and identically distributed (i.i.d.) Gaussian random variables with variance $\frac{1}{N_S}$ and $N_S \geq N_T$. It follows that  $\mathbb{E}(\mathbf{S}\mathbf{S}^\HH)=\mathbf{I}$. 
	%, i.e., $\mathbf{N}_c \sim \mathcal{CN}(\mathbf{0},\sigma_{c}^2\mathbf{I})$. 
	%Note that the sensing and communication are naturally different. The sensing task will be 
	%The communication task is to recover the useful information contained in the transmitted signals $\mathbf{S}$ based on the received signals $\mathbf{Y}_c$, with the knowledge of the channel state information $\mathbf{H}_c$. %During the communication period, $\mathbf{H}_c$ remains stable. Hence, t
	The communication performance can be measured by the MI between the received and transmitted signals, i.e.,
	\begin{equation}\label{MIcomm}
		\begin{split}
			I_c &=  I(\mathbf{Y}_c;\mathbf{S}|\mathbf{H}_c) = \log \left\vert \mathbf{I}+\sigma_{c}^{2}\mathbf{H}_c\mathbf{F}\mathbf{F}^\HH \mathbf{H}_c^\HH \right\vert.
		\end{split}
	\end{equation}

	\subsection{Sensing model}
	The objective for sensing is to estimate the parameters of interest (POI) of targets based on the received signals, with the knowledge of the transmitted probing signal. Assume that the target sensing is performed within a coherent processing interval (CPI) consisting of $N_S$ frames.  %, as illustrated in Fig. \ref{fig_sys}.
	%	\begin{figure}[!t]
		%		\centering
		%		\includegraphics[width=3.0in]{PMN-TMT-2.pdf}
		%		\caption{Illustration of the considered ISAC system. }
		%		\label{fig_sys}
		%	\end{figure}
	%	
	The baseband signal received by the sensing receiver in a CPI is given by \cite{tang2018spectrally}
	\begin{equation}
		\begin{split}
			\mathbf{Y}_s =  \mathbf{H}_s \mathbf{X}+ \mathbf{N}_s \in \mathbb{C}^{N_R\times N_S},
		\end{split}
	\end{equation}
	where $\mathbf{H}_s$ represents the target response matrix, and $\mathbf{N}_s \in \mathbb{C}^{N_R\times N_S}$ is the additive white Gaussian noise. 
	In this paper, we assume  $\mathbf{X}$ is random but known to the sensing receiver. 
	By stacking the columns of $\mathbf{Y}_s^\HH$, we can obtain
	\begin{equation}\label{sigmod_0}
		\begin{split}
			\mathbf{y}_s \triangleq \text{vec} (\mathbf{Y}_s^\HH)= \left(\mathbf{I}_{N_R} \otimes \mathbf{X}\right)^\HH \mathbf{h}_s + \mathbf{n}_s \in \mathbb{C}^{N\times 1},
		\end{split}
	\end{equation}
	where $N = N_TN_R$, $\mathbf{h}_s= \text{vec}(\mathbf{H}_s^\HH)$, and $\mathbf{n}_s= \text{vec}(\mathbf{N}_s^\HH)\sim\mathcal{CN}(\mathbf{0},\sigma_s^2\mathbf{I}_N)$. 
	The correlation matrix $\mathbf{R}\triangleq \mathbb{E}(\mathbf{h}_s\mathbf{h}_s^\HH)$ is given by $\mathbf{R} = \mathbf{R}_R \otimes \mathbf{R}_{T}$ \cite{1021913,1300860,5432999}, where $\mathbf{R}_R$ and $\mathbf{R}_{T}$ denote the correlation matrices at the receiver and the transmitter, respectively. Note that $\mathrm{rank}(\mathbf{R}_R)\leq K$ and $\mathrm{rank}(\mathbf{R}_T)\leq K$, where $K$ denotes the number of sensing targets. 
		In this paper, we assume $\mathbf{R}_R$ and $\mathbf{R}_{T}$ are available to the transmitter. This assumption is justified in ISAC systems, where the prior knowledge of the targets can be obtained from previous estimation  \cite{herbert2017mmse,tang2018spectrally}.
%	The channel correlation matrix $\mathbf{R}\triangleq \mathbb{E}(\mathbf{h}_s\mathbf{h}_s^\HH)$ can be well approximated by $\mathbf{R} = \mathbf{R}_R \otimes \mathbf{R}_{T}$ \cite{965098,1021913,1300860}, where $\mathbf{R}_R$ and $\mathbf{R}_{T}$ denote the correlation matrices at the receiver and the transmitter, respectively, with $\mathrm{rank}(\mathbf{R}_R)\leq K$ and $\mathrm{rank}(\mathbf{R}_T)\leq K$. %Such a model has been widely applied to the MIMO radar systems \cite{965098,1021913,1300860}.
	%As a result, 

	%\subsection{Sensing and communication task}
	%The sensing and communication task are defined as
	%\begin{enumerate}
	%	\item Sensing task: Estimate the parameters of interest (POI) of the target based on the received signals, with the knowledge of the transmitted probing signal. 
	%	\item Communication task: Recover the useful information contained in the transmitted signals based the received signals at the communication receiver, with the knowledge of the channel state information.
	%\end{enumerate}
	
	%The sensing task is to estimate the parameters of interest (POI) of the target based on the received signals, with the knowledge of the transmitted probing signal. 
	For target detection, the sensing receiver shall estimate the target response vector $\mathbf{h}_s$ based on the received signals $\mathbf{y}_s$, or $\mathbf{H}_s$ based on $\mathbf{Y}_s$, equivalently \cite{tang2018spectrally,ren2023fundamental}.
	Assuming a deterministic sensing signal, previous works focused on the system design by optimizing $\mathbf{S}$ to maximize the sensing mutual information (SMI) between the received signals and the target response \cite{tang2018spectrally,yang2007mimo}. 
	Unfortunately, such methods are no longer valid for ISAC systems when random signals are utilized. To this end, we will evaluate the SMI with random signals in the next section.
	
	\section{Sensing Mutual Information with Random Signals}
	In this section, we %will investigate the SMI with random signals and 
	derive an explicit expression for the SMI with random signals by employing random matrix theory. Based on this derived expression, we will study the influence of $N_S$ and investigate the relation between SMI and other sensing performance metrics, such as the ELMMSE and EBCRB.

	%In this section, we will study the SMI with random signals. To give more insights, we give an explicit expression for the SMI based on the random matrix theory. Based on the expression, we study the effect of some important parameters and the relation between the SMI with other sensing performance metrics, e.g., the ergodic LMMSE and Bayesian CRB. 
	Define $\bm{\eta}$ as the POI of the targets, e.g.,
	AOA, AOD and reflection coefficients. We assume that $\bm{\eta}$ remains constant in one CPI and varies between CPIs in an i.i.d. manner. 
	Any estimator for $\bm{\eta}$ aims to estimate a particular realization of $\bm{\eta}$ within one CPI. When $\mathbf{H}_s$ is an injective map of $\bm{\eta}$, the MI between $\bm{\eta}$ and the received signals $\mathbf{y}_s$ is equivalent to that between $\mathbf{h}_s$ and $\mathbf{y}_s$, i.e., $I\left(\bm{\eta};\mathbf{y}_s|\mathbf{S}\right)=I\left(\mathbf{h}_s;\mathbf{y}_s|\mathbf{S}\right)$ \cite{liu2023deterministic}. 
	
	%In this section, we find an explicit expression for SMI in the case that $\mathbf{h}_s$ is assumed to follow a Gaussian distribution, i.e., $\mathbf{h}_s \sim \mathcal{CN}(\mathbf{0},\mathbf{R})$. Such an assumption has been widely utilized for MIMO channel matrix \cite{1300860} and target response matrix \cite{tang2018spectrally,yang2007mimo,liu2023deterministic}. 
	For ease of evaluation, we assume that $\mathbf{h}_s$ follows a Gaussian distribution, i.e., $\mathbf{h}_s \sim \mathcal{CN}(\mathbf{0},\mathbf{R})$, which has been widely adopted for MIMO channel \cite{1300860} and target response matrix \cite{tang2018spectrally,yang2007mimo,liu2023deterministic}. 
	Under such circumstances, the SMI with random signals is defined as \cite{liu2023deterministic}
	\begin{equation}\label{SEMIdef}
		\begin{split}
			%&I\left(\bm{\eta};\mathbf{y}_s|\mathbf{S}\right)\\
			&I_s %= I\left(\bm{\eta};\mathbf{y}_s|\mathbf{S}\right)
			=I\left(\mathbf{h}_s;\mathbf{y}_s|\mathbf{S}\right)\\
			&\triangleq %\mathbb{E}_{\mathbf{S}}\log \left\vert\mathbf{I}+ \sigma_s^{-2}\left(\mathbf{I}_{N_R} \otimes \mathbf{X}\right)^\HH\mathbf{R}\left(\mathbf{I}_{N_R} \otimes \mathbf{X}\right) \right\vert\\
			%&=
			\mathbb{E}_{\mathbf{S}}\log \left\vert\mathbf{I}+ \sigma_s^{-2}\left(\mathbf{R}_R \otimes \mathbf{R}_T^{\frac{1}{2}}\mathbf{F}\mathbf{S}\mathbf{S}^\HH\mathbf{F}^\HH\mathbf{R}_T^{\frac{1}{2}}\right) \right\vert,
		\end{split}
	\end{equation}
	where the expectation is taken over the random signals $\mathbf{S}$. 
	%\footnote{In this paper, we focus on the Gaussian signals, which are widely used in communication. The case with non-Gaussian signals will be left as a future direction.}
	%Note that $I\left(\bm{\eta};\mathbf{y}_s|\mathbf{S}\right)$ represents the SMI that can be achieved. 
	According to \cite{liu2023deterministic}, SMI is directly related to the distortion metric (e.g., the estimation error) of $\bm{\eta}$. 
	%$I\left(\bm{\eta};\mathbf{y}_s|\mathbf{S}\right) = \max_{p(\mathbf{h}_s)}I\left(\bm{\eta};\mathbf{y}_s|\mathbf{S}\right)$, where $p(\mathbf{h}_s)$ denotes the probability density function of $\mathbf{h}_s$. 
	%%While the communication MI between the observation and the transmit data represents the maximum achievable transmission rate, SMI has its own operational meaning. Particularly, 
	%Define $D(R)$ as the distortion-rate function for $\bm{\eta}$, which is monotonically decreasing with $R$. Then, according to \cite{liu2023deterministic}, $D(I\left(\bm{\eta};\mathbf{y}_s|\mathbf{S}\right))$ serves as a lower bound of the average distortion of $\bm{\eta}$. It indicates that $D(I\left(\bm{\eta};\mathbf{y}_s|\mathbf{S}\right))$ represents the minimum average distortion of $\bm{\eta}$.
	%Remarkably, the SMI is monotonically decreasing with the minimum distortion of $\bm{\eta}$ \cite{liu2023deterministic}. 
	%SMI serves as a universal lower bound for sensing	distortion metrics \cite{liu2023deterministic}.
	%where 
	%\begin{equation}\label{SEMIdefcondS}
	%	\begin{split}
		%	I\left(\mathbf{h}_s;\mathbf{y}_s|\mathbf{S}\right)=\log \det\left(\mathbf{I}+ \sigma_s^{-2}\left(\mathbf{I}_{N_R} \otimes \mathbf{X}\right)\mathbf{R} \mathring{\mathbf{X}} \right),
		%	\end{split}
	%\end{equation}
	Unfortunately, the expectation in \eqref{SEMIdef} is computationally prohibitive due to the high-dimensional integrals.

	\subsection{Upper bound of SMI}
	An upper bound of SMI can be obtained by applying the Jensen's inequality  on \eqref{SEMIdef}. In particular, we have 
	%replacing the sample covariance matrix of $\mathbf{S}\mathbf{S}^\HH$ as the statistical covariance matrix $\mathbb{E}(\mathbf{S}\mathbf{S}^\HH)=\mathbf{I}$, i.e., 
	\begin{equation}
		\begin{split}
			I_s\leq\log \left\vert\mathbf{I}+ \sigma_s^{-2}\left(\mathbf{R}_R \otimes \mathbf{R}_T^{\frac{1}{2}}\mathbf{F}\mathbb{E}_{\mathbf{S}}(\mathbf{S}\mathbf{S}^\HH)\mathbf{F}^\HH\mathbf{R}_T^{\frac{1}{2}}\right) \right\vert, 
		\end{split}
	\end{equation}
	where the equality holds when $\mathbf{S}\mathbf{S}^\HH = \mathbf{I}$.
	Accordingly, we can obtain the upper bound as
	\begin{equation}\label{ESEMIdefcondS_Jen}
		\begin{split}
			&I_{s,\max}
			%&\triangleq\log \det\left(\mathbf{I}+ \sigma_s^{-2}\left(\mathbf{R}_R \otimes (\mathbf{R}_T^{\frac{1}{2}}\mathbf{F}\mathbb{E}_{\mathbf{S}}(\mathbf{S}\mathbf{S}^\HH)\mathbf{F}^\HH\mathbf{R}_T^{\frac{1}{2}})\right) \right)\\
			\triangleq\log \left\vert\mathbf{I}+ \sigma_s^{-2} \left(\mathbf{R}_R \otimes \mathbf{R}_T^{\frac{1}{2}}\mathbf{\Phi}\mathbf{R}_T^{\frac{1}{2}}\right) \right\vert,\\
		\end{split}
	\end{equation}
	where $\mathbf{\Phi} = \mathbf{F}\mathbf{F}^\HH$. 
	%\begin{equation}
	%	\begin{split}
		%\mathbf{\Phi} = \mathbf{F}\mathbf{F}^\HH.
		%\end{split}
		%\end{equation}
		Thus, there are two ways to interpret $I_{s,\max}$. 
		%$I_{s,\max}$ has two distinct interpretations.  
		On the one hand, $I_{s,\max}$ can be regarded as the MI obtained by a deterministic signal $\mathbf{S}$ with $\mathbf{S}\mathbf{S}^\HH=\mathbf{I}$. 
		On the other hand, as an upper bound,  $I_{s,\max}$ can be utilized as an approximation for SMI when $N_S$ is large. This is because, as $N_S \to \infty$, the sample covariance matrix $\mathbf{S}\mathbf{S}^\HH$ will tend to be deterministic, i.e., $\lim_{N_S\to \infty} \mathbf{S}\mathbf{S}^\HH \to \mathbf{I}$. Under such circumstances, $I_s$ approaches its upper bound $I_{s,\max}$. 
		%Hence, the upper bound $I_{s,\max}$ can be utilized as an approximation for SMI, especially when $N_S$ is large. 
		Unfortunately, as we will show later, the precision of this approximation is poor when $N_S$ is small.

		\subsection{Asymptotic approximation of SMI}
		In this paper, we derive an explicit expression for the SMI by analyzing the asymptotic behavior of  $I\left(\mathbf{h}_s;\mathbf{y}_s\right)$. In particular, we provide an approximation for the SMI in the large $N_S$ regime based on the first order approximation of the MI.  The result is shown in the following proposition.
		
		%However, the expectation in \eqref{SEMIdef} faces the challenges caused by the high-dimensional integrals, which is computationally prohibitive. Moreover, an explicit expression is useful for performance analysis and system design.
		%To address the above-mentioned issue, we use derive a tractable expression for the SEMI in the following proposition by analyzing the asymptotic behavior of $I\left(\mathbf{h}_s;\mathbf{y}_s|\mathbf{S}\right)$.

		\begin{proposition}\label{TheoRD}
			As $N_S,N_T \to \infty$ with a finite constant ratio $c$, i.e., $N_S/N_T = c$, we have
			%As $N_S\to \infty$, we have
			\begin{equation}\label{SEMI0}
				\begin{split} 
					I_s =  \sum_{j=1}^{K} \bar{\varrho}_{j}(\mathbf{\Phi}) + \mathcal{O}\left(\frac{1}{N_S}\right),
				\end{split}
			\end{equation}
			%	where 
			%	\begin{equation}
				%		\begin{split}	
					%			\bar{\varrho}_{j}=&-\log \det\left(\mathbf{I}_{N_R}+\frac{\lambda_{R,j}}{1+\lambda_{R,j}\delta(\lambda_{R,j})} \mathbf{\Lambda_{T}} \right) \\
					%			&+N_T   \frac{\lambda_{R,j}\delta(\lambda_{R,j})}{1+\lambda_{R,j}\delta(\lambda_{R,j})},
					%		\end{split}	
				%	\end{equation}
			%	$\delta(\rho)$ is the solution to the following equation
			%	\begin{equation}\label{fixeq1}
				%		\delta(\rho) = \frac{1}{N_T} \tr \mathbf{\Lambda_{T}}\left(\mathbf{I}_{N_R}+\frac{\rho}{1+\rho\delta(\rho)} \mathbf{\Lambda_{T}}\right)^{-1},
				%	\end{equation}
			%	and $\mathbf{\Lambda}_{T}=\diag([\lambda_{T,1},\cdots,\lambda_{T,N_T}])$ and $\mathbf{\Lambda}_{R}=\diag([\lambda_{R,1},\cdots,\lambda_{R,N_R}])$ with $\diag(\mathbf{x})$ denoting the diagonal martix whose diagonal is $\mathbf{x}$, and $\{\lambda_{T,i}\}_{i=1}^{N_T}$ and $\{\lambda_{R,j}\}_{j=1}^{N_R}$ denote the eigenvalues of $\mathbf{F}^{\HH}\mathbf{R}_{T}\mathbf{F}$ and $\sigma_s^{-2} \mathbf{R}_{R} $, respectively.
			where 
			\begin{equation}\label{averrhoj0}
				\begin{split}	
					&\bar{\varrho}_{j}(\mathbf{\Phi})=\log \left\vert\mathbf{I}_{N_T}+\frac{\lambda_{R,j}}{1+\lambda_{R,j}\delta(\lambda_{R,j})} \mathbf{T}(\mathbf{\Phi}) \right\vert \\
					&+N_S \log (1+\lambda_{R,j}\delta(\lambda_{R,j}))-N_S   \frac{\lambda_{R,j}\delta(\lambda_{R,j})}{1+\lambda_{R,j}\delta(\lambda_{R,j})},
				\end{split}	
			\end{equation}
			%\begin{equation}\label{Tdef}
			%	\begin{split}
				%		&\mathbf{T}(\mathbf{\Phi})\triangleq \mathbf{R}_{T}^{\frac{1}{2}}\mathbf{F}\mathbf{F}^\HH\mathbf{R}_{T}^{\frac{1}{2}},
				%	\end{split}
			%\end{equation}
			with 
			\begin{equation}
			\mathbf{T}(\mathbf{\Phi})\triangleq \mathbf{R}_{T}^{\frac{1}{2}}\mathbf{\Phi}\mathbf{R}_{T}^{\frac{1}{2}}.
		\end{equation}  
		$\{\lambda_{R,j}\}_{j=1}^{K}$ denotes the eigenvalues of $\sigma_s^{-2} \mathbf{R}_{R}$, and $\delta(\rho)$ is the solution to the following equation
			\begin{equation}\label{fixeq1}
				\delta(\rho) = \frac{1}{N_S} \tr\left( \mathbf{T}(\mathbf{\Phi})\left(\mathbf{I}_{N_T}+\frac{\rho}{1+\rho\delta(\rho)} \mathbf{T}(\mathbf{\Phi})\right)^{-1}\right).
			\end{equation}
		\end{proposition}
		
		\emph{Proof:} See Appendix \ref{proofTheoRD}.  \hfill $\blacksquare$

		%\begin{remark}
		%	\emph{Proposition \ref{TheoRD}} indicates that the ESMI is determined by several parameters: 1) the number of targets $K$, which indicates the number of the parameters to estimate,
		%	2) the correlation at the receiver $\{\lambda_{R,j}\}_{j=1}^{N_R}$, which affects the receive SNR, 3) the eigenvalue of $\mathbf{T}(\mathbf{\Phi})$, which represents the distribution of the transmit power, and 4) the number of frames $N_S$, which affects the sensing degree of freedom (DoF) as we will show later.
		%\end{remark}
		
		%\begin{enumerate}
		%	\item The number of targets, which affects the number of the parameters to estimate.
		%	\item The correlation at the receiver $\{\lambda_{R,j}\}_{j=1}^{N_R}$, which affects the receive SNR.
		%	\item The eigenvalue of $\mathbf{T}(\mathbf{\Phi})$, which represents the distribution of the transmit power.
		%\end{enumerate} 
		
		%\subsection{Sensing degree of freedom}

		%The ESMI provides a novel perspective and tool for analyzing 
		\emph{Proposition \ref{TheoRD}} provides an explicit expression for the SMI between the POI and the received signals. Note that, different from the upper bound in \eqref{ESEMIdefcondS_Jen}, \emph{Proposition \ref{TheoRD}} aims to approximate SMI by analyzing its asymptotic behavior. As will be shown later, although \eqref{SEMI0} is derived under the condition that $N_S$ and $N_T$ approach infinity, it provides accurate approximation for the SMI even when $N_S$ and $N_T$ are small. 

		\subsection{Effect of the number of frames}
		With the explicit expression of the SMI, we can analyze the impact of key system parameters. 
		%For example, the sensing performance using random signals will be improved as $N_S$ increases. However, this has not been theoretically investigated due to the lack of an explicit expression. 
		To evaluate the effect of $N_S$, we calculate the derivative of the SMI with respect to $N_S$. 
		%the previous works have not provided a theoretical analysis on the impact of $N_S$, due to the lack of an explicit expression for the sensing metric with random signals. To address this gap, we aim to investigate the effect of $N_S$ by utilizing the explicit expression of ESMI. 
		%However, the theoretical analysis on the effect of $N_S$ is still unavailable in the previous works due to the lack of explicit expression for the sensing metric with random signals. To fill this blank, we will study the effect of $N_S$ based on the explicit expression of ESMI. 
		For all $N_S\geq 0$, the derivative of $\bar{\varrho}_{j}(\mathbf{\Phi})$, which is given in \eqref{averrhoj0}, with respect to $N_S$ can be expressed in the form of
		\begin{align}
			&\frac{\partial \bar{\varrho}_{j}}{\partial N_S}=-\frac{\lambda_{R,j}^2 \tr\left[\mathbf{T}\left(\mathbf{I}_{N_T}+\frac{\lambda_{R,j}}{1+\lambda_{R,j}\delta(\lambda_{R,j})} \mathbf{T} \right)^{-1}\right]}{(1+\lambda_{R,j}\delta(\lambda_{R,j}))^2}\cdot \frac{\partial \delta(\lambda_{R,j})}{\partial N_S}\notag \\
			&\quad  +\left(\frac{N_S\lambda_{R,j}}{1+\lambda_{R,j}\delta(\lambda_{R,j})}-\frac{N_S\lambda_{R,j}}{(1+\lambda_{R,j}\delta(\lambda_{R,j}))^2}\right)\cdot \frac{\partial \delta(\lambda_{R,j})}{\partial N_S}\notag\\
			&\quad 
			+\log (1+\lambda_{R,j}\delta(\lambda_{R,j}))- \frac{\lambda_{R,j}\delta(\lambda_{R,j})}{1+\lambda_{R,j}\delta(\lambda_{R,j})} \\
			&\overset{(a)}{=}\log (1+\lambda_{R,j}\delta(\lambda_{R,j}))- \frac{\lambda_{R,j}\delta(\lambda_{R,j})}{1+\lambda_{R,j}\delta(\lambda_{R,j})} \geq 0,\notag
		\end{align}
		where step (a) follows \eqref{fixeq1}. Thus, we have \begin{equation}\label{gradientNSall}
			\frac{\partial I_s}{\partial N_S}= \sum_{j=1}^K\frac{\partial \bar{\varrho}_{j}}{\partial N_S}\geq 0.
		\end{equation}
		This indicates that the SMI $I_s$ is monotonically increasing with respect to $N_S$.
		
		\subsection{Sensing DoF loss}
		%To further quantify the effect of $N_S$, 
		To measure the impact of random signals on sensing, we further investigate the sensing Degree of Freedom (DoF). This concept was introduced in \cite{xiong2023fundamental} to measure the loss of CRB induced by the random signals and defined as the effective number of independent observations.
		%Intuitively, the sensing DoF should be defined as the number of ``effective'' independent observations \cite{xiong2023fundamental}. 
		By following the same idea, we define the SMI-oriented sensing DoF as
		\begin{equation}\label{sdofdef}
			\nu_s =  \lim_{\sigma_s^2\to 0}  N_S \frac{ I_s}{I_{s,\max}}.
		\end{equation}
		where the maximum sensing DoF is the total number of the independent observations $N_S$.
		The normalization coefficient $\frac{I_s}{I_{s,\max}}$ is the ratio between the SMI with random signals and its upper bound achieved by deterministic signals. Thus, this ratio measures the MI loss caused by the randomness of signals. The following proposition provides the lower bound for $I\left(\bm{\eta};\mathbf{y}_s|\mathbf{S}\right)$, based on which the maximum sensing DoF loss can be obtained. 
		%To find the maximum of the sensing DoF loss, we provide a lower bound for $I\left(\mathbf{h}_s;\mathbf{y}_s\right)$ in the following proposition.
		\begin{proposition}\label{TheoIlb}
			When $N_S \geq N_T$, we have
			\begin{equation}\label{Ilower bound}
				\begin{split} \! \!  I_s \geq \frac{N_S - \min\{K,\mathrm{rank}(\mathbf{\Phi})\}}{N_S} I_{s,\max}.
				\end{split}
			\end{equation}
		\end{proposition}
		
		\emph{Proof:} See Appendix \ref{proofTheoIlb}. \hfill $\blacksquare$
		
		Based on \emph{Proposition \ref{TheoIlb}}, the sensing DoF can be bounded by 
		\begin{equation}\label{SDoF_range}
			N_S - \min\{K,\mathrm{rank}(\mathbf{\Phi})\} \leq \nu_s \leq N_S.
		\end{equation}
		This indicates that $\min\{K,\mathrm{rank}(\mathbf{\Phi})\}$ is the maximum sensing DoF loss defined based on SMI, which coincides with that defined based on CRB \cite{xiong2023fundamental}. 
		%As the value of $K$ increases, the sensing task becomes more onerous due to various factors, e.g., the presence of the interference and the need for higher resolution. This necessitates a large number of samples to effectively estimate POI. 
		%In particular, more samples enables the system to obtain a more accurate estimation of POI. 
		%In particular, more samples provide a higher level of redundancy, enabling the system to average out the random variations and obtain a more reliable estimate of the target's characteristics. 
		It can be observed that the maximum sensing DoF loss will increase with $K$. This is because the system needs to separate the signals associated with different targets, and this task becomes more challenging as the number of targets increases, due to the interference among different targets. 
		%As the value of $K$ increases, the sensing task becomes more onerous. 
		%In particular, the system needs to separate the signals associated with each target from the overall received signals. This task becomes more challenging as the number of targets increases, due to the interference between different targets. Therefore, the maximum sensing DoF loss will increase. 
		As a result, more samples are required to mitigate such a loss. 
		In particular, according to the squeeze theorem, there is no loss when $N_S \to \infty$, because the lower bound of $I_s$ in \eqref{Ilower bound} approaches $I_{s,\max}$. This agrees with the theoretical results in \eqref{gradientNSall}.

		\section{Relation between SMI and Other Sensing Performance Metrics}
		
		%The sensing process can be regarded as a \emph{non-cooperative joint source channel coding} and 
		
		Section III investigated SMI but the operational meaning of SMI remains unclear \cite{liu2023deterministic}. %In particular, for communication, source coding refers to the representation of the source data, while channel coding focuses on the encoding for reliable transmission. Hence, the communication 
		In particular, for communication, MI has a clear operational meaning and is directly related to the achievable data rate. 
		However, for sensing, the information of the concerned physical parameters is ``encoded'' and ``transmitted'' through the target response matrix to the receiver in a passive manner, and the MI between the target response matrix and the received signal does not bear any direct physical meaning. 
		%Therefore, the target serves as a non-cooperative source which does not have ``coding'' capacity and the operational meaning of the SMI remains unclear.  
		%and the channel represents the target response matrix through which the source is transmitted. 
		%Nevertheless, the operational meaning of the SMI remains unclear. 
		
		To fill in this gap, we will investigate the relation between SMI and other sensing performance metrics, e.g., sensing ELMMSE and Bayesian CRB. By examining these connections, we aim to shed light on the unification of the performance metrics for communication and sensing. For that purpose, we first evaluate the EMMSE, ELMMSE, and Bayesian CRB with random signals. 
		
		\subsection{ELMMSE and EMMSE for estimating $\mathbf{h}_s$}
		To evaluate the average sensing performance with random signals, the ergodic linear minimum mean square error (ELMMSE) was investigated in \cite{lu2023sensing}. %But different from \cite{lu2023sensing} which considers the ELMMSE for estimating $\mathbf{H}_s$, we consider the ELMMSE for estimating $\mathbf{h}_s$ in this paper, i.e.,
		The ELMMSE for estimating $\mathbf{h}_s$ is given by
		\begin{equation}\label{SEMMSEdef}
			\begin{split}
				J_{\mathrm{ELMMSE}} %& = \mathbb{E}_{\mathbf{S}}\; \tr \left(\mathbf{R}^{-1} +\sigma_s^{-2}\mathbf{I}_{N_R} \otimes \mathbf{X}\mathbf{X}^\HH\right)^{-1} \\
				& = \mathbb{E}_{\mathbf{S}}[\tr  (\mathbf{\Psi}_{\mathrm{LMMSE}}(\mathbf{S}))],
			\end{split}
		\end{equation}
		where 
		\begin{equation}  \label{PsiM0}
			\begin{split}
				\mathbf{\Psi}_{\mathrm{LMMSE}}(\mathbf{S}) &\triangleq \mathbb{E} [(\hat{\mathbf{h}}_{s,\text{LMMSE}} - \mathbf{h}_s)(\hat{\mathbf{h}}_{s,\text{LMMSE}} - \mathbf{h}_s)^\HH],
			\end{split}
		\end{equation}
		denotes the data-dependent LMMSE matrix with $\hat{\mathbf{h}}_{s,\text{LMMSE}}$ representing the LMMSE estimate of $\mathbf{h}_s$. 
		For a Gaussian-distributed $\mathbf{h}_s$, we have
		\begin{equation} \label{PsiM}
			\begin{split}
				\mathbf{\Psi}_{\mathrm{LMMSE}}(\mathbf{S})=\mathbf{R}\left(\mathbf{I}+ \sigma_s^{-2}\left(\mathbf{R}_R \otimes \mathbf{R}_T^{\frac{1}{2}}\mathbf{F}\mathbf{S}\mathbf{S}^\HH\mathbf{F}^\HH\mathbf{R}_T^{\frac{1}{2}}\right)\right)^{-1}.
			\end{split}
		\end{equation}
		A lower bound for sensing ELMMSE can be obtained by applying the Jensen's inequality on \eqref{SEMMSEdef}, i.e.,
		\begin{equation}\label{SEMMSEdef2}
			\begin{split}
				J_{\mathrm{ELMMSE}} \geq  \tr \left[ \mathbf{R}\left(\mathbf{I}+ \sigma_s^{-2}\left(\mathbf{R}_R \otimes \mathbf{R}_T^{\frac{1}{2}}\mathbf{\Phi}\mathbf{R}_T^{\frac{1}{2}}\right)\right)^{-1} \right].
			\end{split}
		\end{equation}
		
		%\subsection{Asymptotic approximation of sensing ELMMSE}
		However, an explicit expression for the sensing ELMMSE is not yet available in the literature. In the following, we derive an explicit expression for the sensing ELMMSE by analyzing its asymptotic approximation in the following proposition.
		\begin{proposition}\label{TheoRD2}
			As $N_S,N_T \to \infty$ with a finite constant ratio $c$, i.e., $N_S/N_T = c$, we have
			\begin{equation}\label{SEMMSEdefSE}
				\begin{split}
					&J_{\mathrm{ELMMSE}}= \sum_{j=1}^K   \bar{\varkappa}_{j}(\mathbf{\Phi}) +  \mathcal{O}\left(\frac{1}{N_S}\right),
				\end{split}
			\end{equation}
			where 
			\begin{equation}
				\begin{split}
					\bar{\varkappa}_{j}(\mathbf{\Phi}) = \sigma_s^2\lambda_{R,j} \tr \left(\mathbf{R}_T \mathbf{M}_{T,j}\right),
				\end{split}
			\end{equation}
			with 
			\begin{equation}
				\begin{split}
					\mathbf{M}_{T,j}=\left(\mathbf{I}_{N_T}+\frac{\lambda_{R,j}}{1+\lambda_{R,j}\delta(\lambda_{R,j})} \mathbf{T}(\mathbf{\Phi})\right)^{-1}.
				\end{split}
			\end{equation}
			
		\end{proposition}

		\emph{Proof:} See Appendix \ref{proofTheoRD2}.
		
		\emph{Proposition \ref{TheoRD2}} provides an explicit expression for the ELMMSE of estimating $\mathbf{h}_s$ by analyzing its asymptotic behavior. 
		%Similarly, \emph{Proposition \ref{TheoRD2}} provides an approximation for ELMMSE by analyzing its asymptotic behavior. 
		As will be shown later, \eqref{SEMMSEdefSE} is more accurate than the lower bound in \eqref{SEMMSEdef2}, even when $N_S$ and $N_T$ are small.
		
		%It can be utilized to reveal the operational meaning of maximizing SMI, which is summerized in the following proposition. 
		%With \eqref{SEMMSEdefSE}, we can give the following proposition.
		
		To evaluate the effect of $N_S$, we calculate the derivative for the ELMMSE with respect to $N_S$. 
		For all $N_S\geq 0$, we have
		\begin{equation}\label{deltaNS}
			\begin{split}
				\frac{\partial \delta(\lambda_{R,j})}{\partial N_S} = &- \frac{1}{N_S^2}\tr \left(\mathbf{T}\mathbf{M}_{T,j}\right)\\
				&+\frac{\alpha(\lambda_{R,j})^2\tr \left(\mathbf{M}_{T,j}\mathbf{T}\right)^2}{N_S} \cdot \frac{\partial \delta(\lambda_{R,j})}{\partial N_S},
			\end{split}
		\end{equation}
		where
		\begin{equation}
			\begin{split}
				\alpha(\rho) =\frac{\rho}{1+\rho\delta(\rho)}.
			\end{split}
		\end{equation}
		By solving the linear equation \eqref{deltaNS}, we have
		\begin{equation}
			\begin{split}
				\frac{\partial \delta(\lambda_{R,j})}{\partial N_S}=\frac{-\delta(\lambda_{R,j})}{N_S-\alpha(\lambda_{R,j})^2 \tr \left(\mathbf{T}\mathbf{M}_{T,j}\right)^2}.
			\end{split}
		\end{equation}
		Since $\alpha(\lambda_{R,j})\geq 0$, we can obtain
		\begin{align}
			&\alpha(\lambda_{R,j})^2 \tr \left(\mathbf{T}\mathbf{M}_{T,j}\right)^2= \tr \left(\mathbf{T}\left(\alpha^{-1}(\lambda_{R,j})\mathbf{I}+\mathbf{T}\right)^{-1}\right)^2\nonumber\\
			& = \sum_{i=1}^{N_T} \left(\frac{\lambda_{T,i}}{\alpha^{-1}(\lambda_{R,j})+ \lambda_{T,i}}\right)^2 \leq \mathrm{rank}(\mathbf{T})\leq N_S.
		\end{align}
		Thus, we have $\frac{\partial \delta(\lambda_{R,j})}{\partial N_S} \leq 0.$
		%\begin{equation}
		%	\begin{split}
			%\frac{\partial \delta(\lambda_{R,j})}{\partial N_S} \leq 0.
			%\end{split}
			%\end{equation}
			The derivative of $\bar{\varkappa}_{j}(\mathbf{\Phi})$ with respect to $N_S$ can then be given by 
			\begin{equation}
				\begin{split}
					\frac{\partial \bar{\varkappa}_{j}(\mathbf{\Phi})}{\partial N_S}=\alpha(\lambda_{R,j})^2 \tr \left(\mathbf{M}_{T,j}\mathbf{R}_T\mathbf{M}_{T,j}\mathbf{T}\right) \frac{\partial \delta(\lambda_{R,j})}{\partial N_S}.
				\end{split}
			\end{equation}
			Since $\alpha(\lambda_{R,j})^2 \tr \left(\mathbf{M}_{T,j}\mathbf{R}_T\mathbf{M}_{T,j}\mathbf{T}\right)\geq 0$, we have 
			\begin{equation}
				\begin{split}
					\frac{\partial J_{\mathrm{ELMMSE}}}{\partial N_S} = \sum_{j=1}^K  \frac{\partial \bar{\varkappa}_{j}(\mathbf{\Phi})}{\partial N_S}\leq 0.
				\end{split}
			\end{equation}
			This indicates that the sensing ELMMSE $J_{\mathrm{ELMMSE}}$ is monotonically decreasing with respect to $N_S$.
			
			Given $\mathbf{S}$, the MMSE matrix for estimating $\mathbf{h}_s$ is defined as
			\begin{equation}\label{MMSE}
				\mathbf{\Psi}_{\mathrm{MMSE}}(\mathbf{S})=\mathbb{E}[(\mathbf{h}_s - \mathbb{E}(\mathbf{h}_s|\mathbf{y}_s))(\mathbf{h}_s - \mathbb{E}(\mathbf{h}_s|\mathbf{y}_s))^\HH].
			\end{equation}
			For a Gaussian distributed $\mathbf{h}_s$, the MMSE estimator is identical to the LMMSE estimator. Therefore, we can also define the EMMSE as $J_{\mathrm{EMMSE}} = J_{\mathrm{ELMMSE}}$.

			\subsection{Ergodic Bayesian CRB for estimating $\mathbf{h}_s$}
			The Bayesian CRB provides a lower bound for the MSE of unbiased Bayesian estimators. 
			Different from the conventional CRB, the Bayesian CRB takes the prior information regarding the channel $\mathbf{h}_s$ into account. 
			Based on the model defined in \eqref{sigmod_0}, the Bayesian Fisher information matrix for estimating $\mathbf{h}_s$ is given by  \cite{xiong2023fundamental}
%			\begin{equation}\nonumber
%				\begin{split}
%					&\mathbf{J} \triangleq \mathbb{E}\left( \frac{\partial p_{\mathbf{h}_s}(\mathbf{h}_s)}{\partial \mathbf{h}_s} \frac{\partial p_{\mathbf{h}_s}(\mathbf{h}_s)}{\partial \mathbf{h}_s^\HH}\right)\\
%					&+\mathbb{E} \left(\left.\frac{\partial \log p_{\mathbf{y}_s|\mathbf{S},\mathbf{h}_s}(\mathbf{y}_s|\mathbf{S},\mathbf{h}_s)}{\partial \mathbf{h}_s}\frac{\partial \log p_{\mathbf{y}_s|\mathbf{S},\mathbf{h}_s}(\mathbf{y}_s|\mathbf{S},\mathbf{h}_s)}{\partial \mathbf{h}_s^\HH}\right\vert\mathbf{S}\right) .\\
%				\end{split}
%			\end{equation}
				\begin{align}
					&\mathbf{J} \triangleq \mathbb{E}\left( \frac{\partial p_{\mathbf{h}_s}(\mathbf{h}_s)}{\partial \mathbf{h}_s} \frac{\partial p_{\mathbf{h}_s}(\mathbf{h}_s)}{\partial \mathbf{h}_s^\HH}\right)\\
					&+\mathbb{E} \left(\left.\frac{\partial \log p_{\mathbf{y}_s|\mathbf{S},\mathbf{h}_s}(\mathbf{y}_s|\mathbf{S},\mathbf{h}_s)}{\partial \mathbf{h}_s}\frac{\partial \log p_{\mathbf{y}_s|\mathbf{S},\mathbf{h}_s}(\mathbf{y}_s|\mathbf{S},\mathbf{h}_s)}{\partial \mathbf{h}_s^\HH}\right\vert\mathbf{S}\right) .\notag
			\end{align}
			For a Gaussian-distributed $\mathbf{h}_s$, the BCRB matrix is defined as the inverse of the Bayesian Fisher information matrix, i.e., 
			\begin{equation}\label{CRBmat}
				\begin{split}
					\mathbf{\Psi}_{\mathrm{BCRB}}(\mathbf{S})\triangleq\mathbf{J}^{-1}=\left(\mathbf{R}^{-1} + \sigma_s^{-2}   \mathbf{I}_{N_R} \otimes \mathbf{X}\mathbf{X}^\HH\right)^{-1}.\\
				\end{split}
			\end{equation}
			%\begin{align}
			%			&\mathbf{J}\notag \\
			%	&= \mathbb{E} \left(\left.\frac{\partial \log p_{\mathbf{y}_s|\mathbf{S},\mathbf{h}_s}(\mathbf{y}_s|\mathbf{S},\mathbf{h}_s)}{\partial \mathbf{h}_s}\frac{\partial \log p_{\mathbf{y}_s|\mathbf{S},\mathbf{h}_s}(\mathbf{y}_s|\mathbf{S},\mathbf{h}_s)}{\partial \mathbf{h}_s^\HH}\right\vert\mathbf{S}\right) \notag \\
			%	&\quad + \mathbb{E}\left( \frac{\partial p_{\mathbf{h}_s}(\mathbf{h}_s)}{\partial \mathbf{h}_s} \frac{\partial p_{\mathbf{h}_s}(\mathbf{h}_s)}{\partial \mathbf{h}_s^\HH}\right)\\
			%	&=\sigma_s^{-2} \mathring{\mathbf{X}}  \left(\mathbf{I}_{N_R} \otimes \mathbf{X}\right) +\mathbf{R}^{-1}.  \notag
			%\end{align}
			%
			%\begin{equation}\label{CRBmat}
			%	\begin{split}
				%		\mathbf{\Psi}_{\mathrm{BCRB}} = \mathbf{J}^{-1}.
				%	\end{split}
			%\end{equation}
			To evaluate the sensing performance, specific functions of the BCRB matrix  are normally adopted as the metric of sensing performance, e.g., the determinant \cite{325008} and the trace \cite{yan2015simultaneous}. 
			Hence, we can give the explicit expressions for two forms of EBCRB with random signals based on \emph{Proposition \ref{TheoRD}} and \emph{Proposition \ref{TheoRD2}} as
			%\begin{subequations}
			%\begin{equation}\label{BCRB2}
			%	\begin{split}
				%		\mathrm{EBCRB}_{1} \triangleq \mathbb{E}_{\mathbf{S}}\;\log |\mathbf{\Psi}_{\mathrm{BCRB}}| \approx \sum_{j=1}^{K} \bar{\varrho}_{j}(\mathbf{\Phi}) ,
				%	\end{split}
			%\end{equation}
			%\begin{equation}\label{BCRB1}
			%	\begin{split}
				%			\mathrm{EBCRB}_{2} \triangleq \mathbb{E}_{\mathbf{S}}\;\tr (\mathbf{\Psi}_{\mathrm{BCRB}}) \approx  \sum_{j=1}^K   \bar{\varkappa}_{j}(\mathbf{\Phi}) .
				%		\end{split}
			%\end{equation}
			%\end{subequations}
			\begin{align}
				&\mathrm{EBCRB}_{1} \triangleq \mathbb{E}_{\mathbf{S}}\log |\mathbf{\Psi}_{\mathrm{BCRB}}(\mathbf{S})| \approx \sum_{j=1}^{K} \bar{\varrho}_{j}(\mathbf{\Phi}) ,\label{BCRB2}\\
				&\mathrm{EBCRB}_{2} \triangleq \mathbb{E}_{\mathbf{S}}\tr (\mathbf{\Psi}_{\mathrm{BCRB}}(\mathbf{S})) \approx  \sum_{j=1}^K   \bar{\varkappa}_{j}(\mathbf{\Phi}) \label{BCRB1}.
			\end{align}
			
			%To evaluate the sensing performance, the trace of the CRB matrix is widely adopted as the metric, i.e.,
			%\begin{equation}
			%	\begin{split}
				%		\mathrm{CRB}_{\mathbf{h}_s} = \tr \mathbf{\Psi}_{\mathrm{BCRB}}.
				%	\end{split}
			%\end{equation}

			\subsection{Relation between different metrics}
	 For estimating a Gaussian-distributed $\mathbf{h}_s$, the following proposition reveals the relations between SMI, EBCRB, EMMSE, and ELMMSE. 
			\begin{proposition} \label{PropCRB_MI_MMSE}
				For a Gaussian-distributed $\mathbf{h}_s$, the SMI $I\left(\mathbf{h}_s;\mathbf{y}_s|\mathbf{S}\right)$ defined in \eqref{SEMIdef}, the MMSE matrix $\mathbf{\Psi}_{\mathrm{MMSE}}$ defined in \eqref{MMSE}, the LMMSE matrix $\mathbf{\Psi}_{\mathrm{LMMSE}}$ defined in \eqref{PsiM}, and the BCRB matrix $\mathbf{\Psi}_{\mathrm{BCRB}}$ defined in \eqref{CRBmat} have the following relation 
				\begin{equation}\label{relation3330}
					\begin{split}
						\mathbf{\Psi}_{\mathrm{BCRB}}(\mathbf{S}) = \mathbf{\Psi}_{\mathrm{MMSE}}(\mathbf{S})= \mathbf{\Psi}_{\mathrm{LMMSE}}(\mathbf{S}),
					\end{split}
				\end{equation}
				and
				\begin{equation}\label{relation333}
					\begin{split}
						I\left(\mathbf{h}_s;\mathbf{y}_s|\mathbf{S}\right) =  \log|\mathbf{R}|-\mathbb{E}_{\mathbf{S}}\log |\mathbf{\Psi}_{\mathrm{BCRB}}(\mathbf{S})| .
					\end{split}
				\end{equation}
			\end{proposition}
			
			\emph{Proof:}
			Recalling from  \eqref{PsiM} and \eqref{CRBmat}, we have $\mathbf{\Psi}_{\mathrm{BCRB}}= \mathbf{\Psi}_{\mathrm{LMMSE}}$. Then, according to the squeeze theorem, recalling \eqref{eqIneq1}, we can obtain that $\mathbf{\Psi}_{\mathrm{BCRB}}=\mathbf{\Psi}_{\mathrm{MMSE}}= \mathbf{\Psi}_{\mathrm{LMMSE}}$. 
			
			Recalling \eqref{BCRB2}, we have
			\begin{equation}\label{PsiM11}
				\begin{split}
					&\log|\mathbf{\Psi}_{\mathrm{BCRB}}(\mathbf{S})| \\
					& = \log |\mathbf{R}| - \log \left\vert\mathbf{I}+ \sigma_s^{-2}\left(\mathbf{R}_R \otimes \mathbf{R}_T^{\frac{1}{2}}\mathbf{F}\mathbf{S}\mathbf{S}^\HH\mathbf{F}^\HH\mathbf{R}_T^{\frac{1}{2}}\right)\right\vert.
				\end{split}
			\end{equation}
			By taking the expectation with respect to $\mathbf{S}$ on both sides of \eqref{PsiM11} and recalling \eqref{SEMIdef}, we have
			\begin{equation}\label{PsiM12}
				\begin{split}
					\mathbb{E}_{\mathbf{S}}\log|\mathbf{\Psi}_{\mathrm{BCRB}}(\mathbf{S})|  = \log |\mathbf{R}| - I\left(\mathbf{h}_s;\mathbf{y}_s|\mathbf{S}\right),
				\end{split}
			\end{equation}   
			which completes the proof.  \hfill$\blacksquare$

			According to the definition of MI, we have
			\begin{equation}
				I\left(\mathbf{h}_s;\mathbf{y}_s|\mathbf{S}\right)= H(\mathbf{h}_s|\mathbf{S}) - H(\mathbf{h}_s|\mathbf{y}_s,\mathbf{S}),
			\end{equation} 
			where $H(\cdot)$ denotes the entropy operator. Since the target response matrix $\mathbf{h}_s$ follows the Gaussian distribution $\mathcal{CN}(\mathbf{0},\mathbf{R})$ and is independent of $\mathbf{S}$, we have $H(\mathbf{h}_s|\mathbf{S}) = H(\mathbf{h}_s) =  \log |\mathbf{R}|.$ 
			%\begin{equation}
			%	H(\mathbf{h}_s|\mathbf{S}) = H(\mathbf{h}_s) =  \log |\mathbf{R}|.
			%\end{equation} 
			Then, we have
			\begin{equation}
				H(\mathbf{h}_s|\mathbf{y}_s,\mathbf{S}) = \log|\mathbf{R}| - I\left(\mathbf{h}_s;\mathbf{y}_s|\mathbf{S}\right).
			\end{equation}
			%It follows from \eqref{relation333} that 
			%It indicates that \eqref{relation333} can be reformulated as
			It follows from  \eqref{PsiM} that
			\begin{equation}\label{relation3332}
				\begin{split}
					H(\mathbf{h}_s|\mathbf{y}_s,\mathbf{S}) = \mathbb{E}_{\mathbf{S}}\log |\mathbf{\Psi}_{\mathrm{BCRB}}(\mathbf{S})| .
				\end{split}
			\end{equation}

			\begin{remark}
				\emph{Proposition \ref{PropCRB_MI_MMSE}} 
				%\eqref{relation3332} 
				reveals some insights:
				\begin{enumerate}
					\item EBCRB vs EMMSE/ELMMSE (the connection labeled as \emph{Remark 1.1} in Fig. \ref{illu_connection}):  %According to \cite[Theorem 2]{reeves2018mutual}, for an arbitrarily-distributed $\mathbf{h}_s$, we have $\mathbf{\Psi}_{\mathrm{BCRB}} \preceq \mathbf{\Psi}_{\mathrm{LMMSE}}$. But f
					For the Gaussian-distributed $\mathbf{h}_s$, we know that, the BCRB matrix is identical to the LMMSE and MMSE matrix, i.e., $\mathbf{\Psi}_{\mathrm{BCRB}}(\mathbf{S})=\mathbf{\Psi}_{\mathrm{MMSE}}(\mathbf{S})= \mathbf{\Psi}_{\mathrm{LMMSE}}(\mathbf{S})$.
					%Moreover, according to the squeeze theorem, recalling \eqref{eqIneq1}, we can obtain that $\mathbf{\Psi}_{\mathrm{BCRB}}=\mathbf{\Psi}_{\mathrm{MMSE}}= \mathbf{\Psi}_{\mathrm{LMMSE}}$. 
					This indicates that the estimation error of MMSE and LMMSE estimator can approach BCRB. 
					\item SMI vs EBCRB (the connection labeled as \emph{Remark 1.2} in Fig. \ref{illu_connection}): 
					The equation  $I\left(\mathbf{h}_s;\mathbf{y}_s|\mathbf{S}\right) = \log|\mathbf{R}|- \mathbb{E}_{\mathbf{S}}\log |\mathbf{\Psi}_{\mathrm{BCRB}}(\mathbf{S})| $ indicates that SMI is directly related to Bayesian Fisher information as $\mathbf{\Psi}_{\mathrm{BCRB}} = \mathbf{J}^{-1}$.

					This equivalence offers an interesting connection between information theory and estimation theory. In particular, wireless sensing can be interpreted as ``\emph{non-cooperative}'' communication, where the target plays the role of a non-cooperative transmitter, which encodes information about the POI $\bm{\eta}$ into the target response matrix $\mathbf{h}_s$ and transmits $\mathbf{h}_s$ to the sensing receiver. Here, we use $H(\mathbf{h}_s) =  \log |\mathbf{R}|$ to represent the uncertainty about the target response $\mathbf{h}_s$. 
					Then, the sensing receiver will decode this information to estimate the POI based on the echoes $\mathbf{y}_s$, where $H(\mathbf{h}_s|\mathbf{y}_s,\mathbf{S})$ denotes the uncertain about $\mathbf{h}_s$ when $\mathbf{y}_s$ is known. 
					
					The mutual information $I\left(\mathbf{h}_s;\mathbf{y}_s|\mathbf{S}\right)= H(\mathbf{h}_s|\mathbf{S}) - H(\mathbf{h}_s|\mathbf{y}_s,\mathbf{S})$ is the reduction of the uncertainty about $\mathbf{h}_s$, when $\mathbf{y}_s$ is observed. 
					From the perspective of estimation theory, the utilization of certain estimators to recover POI from $\mathbf{y}_s$ inevitably introduces estimation errors. The MSE of Bayesian estimators with random signals is lower-bounded by the EBCRB, i.e.,   $\mathbb{E}_{\mathbf{S}}\log |\mathbf{\Psi}_{\mathrm{BCRB}}(\mathbf{S})|$. 	
					% \cite{yan2015simultaneous}
					%	\begin{equation}
						%			\mathbb{E} \left(\hat{\mathbf{h}}_s - \mathbf{h}_s\right)\left(\hat{\mathbf{h}}_s - \mathbf{h}_s\right)^\HH\succeq \mathbf{\Psi}_{\mathrm{BCRB}},
						%			%\mathbb{E} \left(\bm{\epsilon}_s\bm{\epsilon}_s^\HH\right) \succeq \mathbf{\Psi}_{\mathrm{BCRB}},
						%		\end{equation}
					%	where $\hat{\mathbf{h}}_s$ denotes any Bayesian estimator of $\mathbf{h}_s$.
					%From estimation theory's point of view, when certain estimators are utilized to recover the POI from $\mathbf{y}_s$, there will always be some estimation error and the MSE is lower bounded by the BCRB $\mathbb{E}_{\mathbf{S}}\log |\mathbf{\Psi}_{\mathrm{BCRB}}|$. 
					The equality $\mathbb{E}_{\mathbf{S}}\log |\mathbf{\Psi}_{\mathrm{BCRB}}(\mathbf{S})|=H(\mathbf{h}_s|\mathbf{y}_s,\mathbf{S})$ indicates that such inevitable errors are identical to the uncertainty of $\mathbf{h}_s$ when $\mathbf{y}_s$ is observed, which comes from the noise or ill-conditioned channel. In other words, the SMI $I\left(\mathbf{h}_s;\mathbf{y}_s|\mathbf{S}\right)$ can be interpreted as \emph{the maximum amount of information} about $\mathbf{h}_s$ that can be extracted by Bayesian estimators.

					\item SMI vs EMMSE/ELMMSE (the connection labeled as \emph{Remark 1.3} in Fig. \ref{illu_connection}): From \emph{Proposition \ref{PropCRB_MI_MMSE}}, we have $I\left(\mathbf{h}_s;\mathbf{y}_s|\mathbf{S}\right) = \log|\mathbf{R}|- \mathbb{E}_{\mathbf{S}}\log |\mathbf{\Psi}_{\mathrm{MMSE}}(\mathbf{S})|$. 
					Given the log-determinant of MMSE matrix has also been widely used to represent MMSE \cite{shi2011iteratively}, we can see that SMI is directly related to EMMSE and ELMMSE. 
				\end{enumerate}
			\end{remark}

			%The above connections between SMI and traditional sensing metrics, e.g., BCRB and ELMMSE, indicate that SMI can serve as an effective performance metric for measuring sensing performance. 
			The aforementioned connections between SMI and traditional sensing metrics, e.g., EBCRB, EMMSE, and ELMMSE, highlight the effectiveness of SMI as a metric for evaluating sensing performance. 
			Given MI has been widely utilized to evaluate communication performance, we propose to leverage MI as a unified metric for both sensing and communication. 
			%for unifying the performance metrics for sensing and communication. 
			%In the following section, we illustrate the application of SMI in the design of ISAC systems.
			
			%how the SMI can be utilized for designing ISAC systems.  
			
			%Moreover, maximizing the SMI is intrinsically equivalent to minimizing the BCRB defined in \eqref{BCRB2}. This equivalence offers an interesting connection between information theory and estimation theory. 
			%In particular, wireless sensing can be interpreted as ``\emph{non-cooperative}'' communication, where the target plays the role of a non-cooperative transmitter, which encodes information about the POI $\eta$ into the target response matrix $\mathbf{h}_s$ and transmits $\mathbf{h}_s$ to the sensing receiver. Then, the sensing receiver will decode this information to estimate the POI based on the echoes $\mathbf{y}_s$. When different estimators of $\mathbf{h}_s$ are utilized to recover the POI, the estimators themselves will introduce some uncertainty and cause estimation error. 

				When $\mathbf{h}_s$ is non-Gaussian distributed, the relation between SMI and other sensing metrics is different. In particular, with a deterministic signal $\mathbf{S}$, the relation between BCRB, MMSE and LMMSE is given by \cite{reeves2018mutual}
				\begin{equation}\label{eqIneq1}
					\mathbf{\Psi}_{\mathrm{BCRB}}(\mathbf{S})\preceq \mathbf{\Psi}_{\mathrm{MMSE}}(\mathbf{S})\preceq \mathbf{\Psi}_{\mathrm{LMMSE}}(\mathbf{S}),
				\end{equation}
				which indicates that, the MMSE is a lower bound of LMMSE. This is because the LMMSE estimator is obtained by minimizing the MSE within linear estimators. Meanwhile, BCRB is a lower bound for the estimation error of Bayesian estimators, including both MMSE and LMMSE estimators. Moreover, given any realization of $\mathbf{S}$, i.e., $\mathbf{S} = \mathbf{A}$, the relation between BCRB and MI is given by \cite{reeves2018mutual}
				\begin{equation}\label{eqIneq2}
					I\left(\mathbf{h}_s;\mathbf{y}_s|\mathbf{S}=\mathbf{A}\right)\geq -\log|\mathbf{\Psi}_{\mathrm{BCRB}}(\mathbf{A})|-\log|\mathbf{J}_X|,
				\end{equation}
				where
				\begin{equation}
					\mathbf{J}_X \triangleq \mathbb{E}\left( \frac{\partial p_{\mathbf{h}_s}(\mathbf{h}_s)}{\partial \mathbf{h}_s} \frac{\partial p_{\mathbf{h}_s}(\mathbf{h}_s)}{\partial \mathbf{h}_s^\HH}\right).
				\end{equation}
				By taking expectation over $\mathbf{A}$ on both sides of \eqref{eqIneq2}, we have
				\begin{equation}\label{eqIneq3}
					I\left(\mathbf{h}_s;\mathbf{y}_s|\mathbf{S}\right)\geq -\mathbb{E}_{\mathbf{A}}\log|\mathbf{\Psi}_{\mathrm{BCRB}}(\mathbf{A})|-\log|\mathbf{J}_X|.
				\end{equation}
%				For a Gaussian distributed $\mathbf{h}_s$, we have $\log|\mathbf{J}_X| = -\log|\mathbf{R}|$, which indicates that the equality in \eqref{eqIneq3} holds for a Gaussian distributed $\mathbf{h}_s$. 
				%the connections between SMI and other sensing metrics for estimating a general $\mathbf{h}_s$, e.g., MMSE and LMMSE, can not be easily obtained at this stage, 
				%However, the explicit expressions for these sensing metrics with random signals are still unavailable, which will be left as the future directions.
				Nevertheless, the explicit expressions of these sensing metrics for estimating a non-Gaussian $\mathbf{h}_s$ with random signals are not yet available, and will be left as a future research direction. 
				%The other connections can not be easily obtained at this stage. Hence, the connections between SMI and other sensing performance metrics
%			\end{remark}

			\section{SMI-Oriented Precoding Design}
			In this section, we will optimize the precoder to maximize the SMI for sensing-only and ISAC scenarios, respectively. In particular, given SMI serves as an effective metric for evaluating sensing performance, we maximize the SMI under a constraint on the transmit power in the sensing-only scenario. 
			In the ISAC scenario, besides the transmit power constraint, we also introduce one constraint on the communication MI. By doing this, we aim to demonstrate that MI can serve as a unified metric for ISAC.

			\subsection{Sensing-only scenario}
			For sensing-only systems, the optimization problem is formulated as
			\begin{equation}\label{P00}
				\begin{array}{ccl}
					&\min\limits_{\mathbf{\Phi}\in \mathcal{A}} &\mathcal{L}_s(\mathbf{\Phi}),\\
				\end{array}
			\end{equation}
			where  
			\begin{equation}
				\mathcal{L}_s(\mathbf{\Phi})=-I\left(\bm{\eta};\mathbf{y}_s|\mathbf{S}\right),
			\end{equation}
			and
			$\mathcal{A} =  \{\mathbf{\Phi} \in \mathbb{C}^{N_T\times N_T}|\mathbf{\Phi}=\mathbf{\Phi}^\HH, \tr(\mathbf{\Phi}) \leq P\}$ denotes the feasible region, which indicates that 
			%The coefficient $\omega\in [0,1]$ denotes the pre-given weight for communication performance. 
			the transmit power is constrained by a maximum value of $P$. 
			%$\mathcal{L}(\mathbf{\Phi})=\sum_{j=1}^{K} \bar{\varrho}_{j}(\mathbf{\Phi})$ and  the transmit power is constrained to a maximum value of $P$. 
			A common practice to solve \eqref{P00} is to utilize the interior point method associated with the Newton's method, which requires both gradient and Hessian matrix. However, since $\delta$ is obtained by solving the equation defined in \eqref{fixeq1},  it is difficult to obtain the Hessian matrix of $\mathcal{L}_s(\mathbf{\Phi})$ with respect to $\mathbf{\Phi}$. To address this issue, we propose to solve the problem in \eqref{P00} by exploiting the gradient-projection (GP) method, which makes it easier to deal with the constraint and only requires the gradient.

			Define $\nabla_{\mathbf{\Phi}}	\mathcal{L}_s(\mathbf{\Phi})$ as the gradient of the objective function in \eqref{P00} with respect to $\mathbf{\Phi}$. 
			Note that $\delta$ defined in \eqref{fixeq1} is dependent on $\mathbf{\Phi}$, which makes the derivative more complex. To this end, we first obtain the derivatives of $\delta(\rho)$ with respect to $\mathbf{\Phi}$by the following lemma.

			\begin{lemma}\label{lemmaderiveRs}
				Define $\alpha(\rho)=\frac{\rho}{1+\rho\delta(\rho)}$. Then, the gradient of $\delta(\rho)$ with respect to $\mathbf{\Phi}$, i.e., $\mathbf{\Delta}_{\rho}'\triangleq
				\frac{\partial \delta(\rho)}{\partial \mathbf{\Phi}^*}$ is obtained by
				\begin{equation}\label{derivaLER}
					\begin{split}
						&\mathbf{\Delta}_{\rho}'  = \frac{\mathbf{R}_T^{\frac{1}{2}}\left(\mathbf{I}+\alpha(\rho) \mathbf{T}(\mathbf{\Phi})\right)^{-2}\mathbf{R}_T^{\frac{1}{2}}}{N_S-\alpha^2(\rho)\tr(\mathbf{T}(\mathbf{\Phi}) \left(\mathbf{I}+\alpha(\rho) \mathbf{T}(\mathbf{\Phi})\right)^{-1})^2 }.
					\end{split}
				\end{equation}
			\end{lemma}
			
			\emph{Proof}: See Appendix \ref{prooflemmaderiveRs}. \hfill $\blacksquare$
			
			It is interesting to observe that, although $\delta$ should be obtained by a fixed-point equation, the explicit expression of its gradient is available. 
			
			Then,  the gradient of $\bar{\varrho}_{j}(\mathbf{\Phi})$ with respect to $\mathbf{\Phi}$ is given in the following proposition.
			\begin{proposition} \label{PropEQ}
				The gradient of $\bar{\varrho}_{j}(\mathbf{\Phi})$ with respect to $\mathbf{\Phi}$ is given by
				\begin{align}\label{DeltarhojF}
					\nabla_{\mathbf{\Phi}}\bar{\varrho}_{j}(\mathbf{\Phi})\triangleq \frac{\partial \bar{\varrho}_{j}(\mathbf{\Phi})}{\partial \mathbf{\Phi}^*}= \alpha_j \mathbf{R}_T^{\frac{1}{2}}\mathbf{M}_{T,j}\mathbf{R}_T^{\frac{1}{2}}.
				\end{align}
			\end{proposition}
			
			\emph{Proof:} See Appendix \ref{proofPropEQ}. \hfill $\blacksquare$
			
			\begin{corollary}\label{PropEQ2}
				The gradient of $\bar{\varkappa}_{j}(\mathbf{\Phi})$ with respect to $\mathbf{\Phi}$ is given by
				\begin{equation}\label{DeltavarejF}
					\begin{split}
						&\nabla_{\mathbf{\Phi}}\bar{\varkappa}_{j}(\mathbf{\Phi})\triangleq  \frac{\partial \bar{\varkappa}_{j}(\mathbf{\Phi})}{\partial \mathbf{\Phi}^*}\\
						&=\sigma_s^2\lambda_{R,j} \alpha_j^2\tr\left(\mathbf{R}_T \mathbf{M}_{T,j}\mathbf{T} \mathbf{M}_{T,j}\right)  \mathbf{\Delta}_{\lambda_{R,j}}'(\mathbf{\Phi})\\
						&\quad-\sigma_s^2\lambda_{R,j}\alpha_j\mathbf{R}_T^{\frac{1}{2}}\mathbf{M}_{T,j}\mathbf{R}_T\mathbf{M}_{T,j}\mathbf{R}_T^{\frac{1}{2}},
					\end{split}
				\end{equation}
				where %$\mathbf{\Delta}_{\lambda_{R,j}}'(\mathbf{\Phi})= \mathbf{\Delta}_{\rho}'|_{\rho=\lambda_{R,j}}$. 
				\begin{equation}
					\mathbf{\Delta}_{\lambda_{R,j}}'(\mathbf{\Phi})= \mathbf{\Delta}_{\rho}'|_{\rho=\lambda_{R,j}}.
				\end{equation}
			\end{corollary}
			\emph{Proof:} See Appendix \ref{proofPropEQ2}. \hfill $\blacksquare$
			
			%Then, the gradient of $\bar{\varrho}_{j}(\mathbf{\Phi})$ with respect to $\mathbf{\Phi}$ is given by
			%\begin{align}\label{DeltarhojF}
			%	\nabla_{\mathbf{\Phi}}\bar{\varrho}_{j}(\mathbf{\Phi})\triangleq \frac{\partial \bar{\varrho}_{j}(\mathbf{\Phi})}{\partial \mathbf{\Phi}^*}= \alpha_j \mathbf{R}_T^{\frac{1}{2}}\mathbf{M}_{T,j}\mathbf{R}_T^{\frac{1}{2}}.
			%\end{align}
			Therefore, $\nabla_{\mathbf{\Phi}}	\mathcal{L}_s(\mathbf{\Phi})$ is computed as 
			%$\nabla_{\mathbf{\Phi}}	\mathcal{L}_s(\mathbf{\Phi}) =-\sum_{j=1}^{K}  \nabla_{\mathbf{\Phi}}\bar{\varrho}_{j}(\mathbf{\Phi})$, 
			\begin{equation}
				\begin{split}
					\nabla_{\mathbf{\Phi}}	\mathcal{L}_s(\mathbf{\Phi}) =-\sum_{j=1}^{K}  \nabla_{\mathbf{\Phi}}\bar{\varrho}_{j}(\mathbf{\Phi}),
				\end{split}
			\end{equation}
			where $\nabla_{\mathbf{\Phi}}\bar{\varrho}_{j}(\mathbf{\Phi})$ is defined in \eqref{DeltarhojF}.
			
			Then, at the $m$th iteration, $\mathbf{\Phi}^{(m+1)}$ is updated by
			\begin{equation}\label{RiemGradF0}
				\begin{split}
					&\widehat{\mathbf{\Phi}}^{(m+1)}=\mathbf{\Phi}^{(m)} -  \beta_{m} \nabla_{\mathbf{\Phi}} \mathcal{L}_s(\mathbf{\Phi}^{(m)}),
				\end{split}
			\end{equation}
			where  $\beta_{m}$ denotes the step size obtained via the Armijo line search \cite[Definition 4.2.2]{absil2009optimization}. A projection is needed to remap the updated points onto the feasible region. The projection of $\mathcal{A}$ is defined as 
			\begin{equation}
				\mathbf{\Phi}^{(m+1)} = \mathcal{F}\left(\widehat{\mathbf{\Phi}}^{(m+1)}\right),
			\end{equation}
			where
			\begin{equation}\label{retractionF0}
				\begin{split}
					\mathcal{F}(\mathbf{\Phi})=\left\{
					\begin{array}{lcl}
						&\mathbf{\Phi},&\mathrm{if \;} \mathbf{\Phi}\in\mathcal{A}\\
						&\frac{P}{\tr (\mathbf{\Phi})}\mathbf{\Phi},& \mathrm{otherwise}
					\end{array}
					\right. .
				\end{split}
			\end{equation}
			%	To solve (\ref{P00}), we propose the GD-based method summarized in \emph{Algorithm \ref{ALG0}}, whose convergence is guaranteed by \cite{boyd2004convex}.
			The convergence of the proposed GP-based method is guaranteed by \cite{boyd2004convex}.
			%	\begin{remark}
				%		Based on \emph{Proposition \ref{TheoRD2}}, the ELMMSE-oriented precoding design problem can be formulated as
				%		\begin{equation}\label{P002}
					%			\begin{array}{ccl}
						%				&\min\limits_{\mathbf{F}} &J_{\mathrm{ELMMSE}}(\mathbf{\Phi})\\
						%				& s.t. & \Vert\mathbf{F}\Vert^2 \leq P.
						%			\end{array}
					%		\end{equation}
				%	By replacing $\nabla_{\mathbf{\Phi}}	\mathcal{L}(\mathbf{\Phi})$ as \eqref{DeltavarejF}, the problem \eqref{P002} can be solved by \textbf{Algorithm \ref{ALG0}}.
				%	\end{remark}
			
			%	\begin{algorithm}[h] 
				%		\caption{Proposed GD-based Method to optimize $\mathbf{\Phi}^{\dagger}$} 
				%		\textbf{Input:} An initial point $\mathbf{\Phi}_{0}$ and $m=0$.
				%		
				%		\textbf{Repeat}
				%		\begin{enumerate} 
					%			\item Compute $\beta_{m}$ via the Armijo line search step.
					%			\item Update $\widehat{\mathbf{\Phi}}^{(m+1)}$ with $\widehat{\mathbf{\Phi}}^{(m)}$ via (\ref{RiemGradF0}).
					%			\item Update $\mathbf{\Phi}^{(m+1)}$ by retracting $\mathbf{\Phi}^{(m+1)}$ via \eqref{retractionF0}.
					%			\item $m \gets m+1$.
					%		\end{enumerate} 
				%		\textbf{Until} Convergence criterion is met.\\
				%		\textbf{Output:} The optimal solution $\mathbf{\Phi}^{\dagger}=\mathbf{\Phi}_{m}$.
				%		\label{ALG0}
				%	\end{algorithm}

			%\section{SMI-Oriented Precoding Design for ISAC}
			\subsection{ISAC scenario}
			Next, we will optimize the precoder to maximize the SMI for ISAC system while guaranteeing the communication performance. The optimization problem is formulated as
			\begin{equation}\label{P0}
				\begin{array}{ccl}
					&\min\limits_{\mathbf{\Phi}\in\mathcal{A}} &\mathcal{L}_s(\mathbf{\Phi})\\
					& s.t. & I_c(\mathbf{\Phi})\geq R_0,
				\end{array}
			\end{equation}
			where $I_c(\mathbf{\Phi}) $ denotes the communication data rate, which is defined in \eqref{MIcomm}. 
			The main difference between \eqref{P0} and  \eqref{P00} is the constraint where  $R_0$ denotes a required level of communication performance. Note that the expression of $\mathcal{L}_s(\mathbf{\Phi})$ is complex with respect to $\mathbf{\Phi}$. %, because $\delta$ depends on $\mathbf{\Phi}$ and can be obtained by solving a linear equation. 
			Therefore, although both the cost function and the constraints are convex, \eqref{P0} is hard to solve directly. 
			%The problem \eqref{P0} is non-convex and is difficult to solve directly. 
			To this end, we exploit the ADMM by introducing an auxiliary variable $\mathbf{\Omega}$, such that $\eqref{P0}$ can be recast as
			\begin{equation}\label{P1}
				\begin{array}{ccl}
					\mathcal{P}_1:&\min\limits_{\mathbf{\Phi},\mathbf{\Omega}\in\mathcal{A}} &\mathcal{L}_s(\mathbf{\Phi})\\
					& s.t. & I_c(\mathbf{\Omega})\geq R_0, \quad \mathbf{\Omega} = \mathbf{\Phi}.
				\end{array}
			\end{equation}
			The problem \eqref{P1} leads to the augmented Lagrangian function
			\begin{equation}
				\mathcal{L}(\mathbf{\Phi},\mathbf{\Omega},\mathbf{\Delta})=\mathcal{L}_s(\mathbf{\Phi}) + \frac{\rho}{2}\left\Vert \mathbf{\Phi} - \mathbf{\Omega} + \mathbf{\Delta} \right\Vert^2,
			\end{equation}
			where $\rho$ is a penalty parameter.
			Then, at the $m$th iteration, the variables can be updated as
			\begin{subequations}\label{ADMMiteration}
				\begin{align}	
					&\mathbf{\mathbf{\Phi}}^{(m+1)} =\arg \min_{\mathbf{\mathbf{\Phi}}\in\mathcal{A}} \mathcal{L}(\mathbf{\Phi},\mathbf{\Omega}^{(m)},\mathbf{\Delta}^{(m)}),\label{subprobu}\\
					&\mathbf{\Omega}^{(m+1)}=\arg \min_{\mathbf{\Omega}\in \mathcal{A}} \mathcal{L}(\mathbf{\Phi}^{(m+1)},\mathbf{\Omega},\mathbf{\Delta}^{(m)}),\label{subprobv}\\
					&\quad \quad \quad \quad \quad \text{s.t.}\;I_c(\mathbf{\Omega})\geq R_0,\notag\\
					&\mathbf{\Delta}^{(m+1)}=\mathbf{\Delta}^{(m)}+\mathbf{\mathbf{\Phi}}^{(m+1)}-\mathbf{\Omega}^{(m+1)},\label{zup}
				\end{align}
			\end{subequations}
			where $\mathbf{\mathbf{\Phi}}^{(m+1)}$, $\mathbf{\Omega}^{(m)}$, and $\mathbf{\Delta}^{(m)}$ denotes
			$\mathbf{\Phi}$, $\mathbf{\Omega}$, and $\mathbf{\Delta}$ at the $m$th iteration, respectively.

			\subsubsection{Update $\mathbf{\mathbf{\Phi}}^{(m+1)}$ via \eqref{subprobu}}
			Since \eqref{subprobu} resembles \eqref{P00}, we propose to solve \eqref{subprobu} by a GP-based method similar to the proposed method in the sensing-only scenario. The gradient of $\mathcal{L}(\mathbf{\Phi}|\mathbf{\Omega}^{(m)},\mathbf{\Delta}^{(m)})$ with respect to $\mathbf{\Phi}$ is 
			\begin{equation}
				\begin{split}
					&\nabla_{\mathbf{\Phi}} \mathcal{L}(\mathbf{\Phi}|\mathbf{\Omega}^{(m)},\mathbf{\Delta}^{(m)}) \\
					&= -\sum_{j=1}^{K}  \nabla_{\mathbf{\Phi}}\bar{\varrho}_{j}(\mathbf{\Phi}) + \rho\left(\mathbf{\mathbf{\Phi}}-\mathbf{\Omega}^{(m)}+\mathbf{\Delta}^{(m)}\right)^\HH\mathbf{\mathbf{\Phi}}.
				\end{split}
			\end{equation}
			Then, $\mathbf{\Phi}^{(m+1)}$ is updated by
			\begin{equation}\label{RiemGradF2}
				\begin{split}
					&\mathbf{\Phi}^{(m+1)}=\mathbf{\Phi}^{(m)} -  \beta_{m} \nabla_{\mathbf{\Phi}} \mathcal{L}(\mathbf{\Phi}^{(m)}).
				\end{split}
			\end{equation}

			\subsubsection{Update $\mathbf{\Omega}^{(m+1)}$ via \eqref{subprobv}}
			The resultant problem \eqref{subprobv} is equivalent to
			\begin{equation}\label{RiemGradF21}
				\begin{split}
					\min_{\mathbf{\Omega}\in \mathcal{A}} \; &\left\Vert \mathbf{\Phi}^{(m+1)} - \mathbf{\Omega} + \mathbf{\Delta}^{(m)} \right\Vert^2\\
					s.t. \; & I_c(\mathbf{\Omega})\geq R_0.
				\end{split}
			\end{equation}
			This problem is a typical semi-definite programming (SDP) problem because the objective function and all constraints are reformulated as a linear or quadratic form. This problem can be easily solved by the CVX toolbox \cite{grant2014cvx}. 
			
			%\xl{Convergence}
			
			The cost function will not increase over the ADMM iteration process given in \eqref{ADMMiteration}. According to the monotone bounded theorem \cite{bibby1974axiomatisations}, the iteration will converge to a set of stationary points in the feasible set.

			\section{Simulation Results}
			In this section, we will validate the accuracy of the theoretical analysis and the effectiveness of proposed optimization-based algorithms by simulation. 
			We consider a mmWave system operating at a carrier frequency of 28 GHz \cite{6834753}. 
			The AOD and AOA of the targets are generated uniformly in the range  $[30^\circ,60^\circ]$. The number of antennas at the transmitter and receiver is set as $N_T = 16$ and $N_R =16$, respectively. 
			%The number of frames in one pulse is $N_s=128$, unless otherwise specified.
			%The noise power is $\sigma^2=-90$ dBm. 
			For the GP-based optimization, we set the maximum number of iterations as $40$. To terminate the iteration, the tolerance for the norm of the gradient between two iterations is set as $10^{-10}$. The transmission power is set as $P = 30$ dBm, the noise power is $\sigma^2=-90$ dBm, and the SNR at the sensing receiver is 10 dB, unless otherwise specified. %We set $\mathbb{E}(||\mathbf{H}_s||_F^2)=10^{-11}$. %the variance of $\epsilon_{k}$ is set as $\sigma_{k}^2 = 10^{-12}$. %The signal-to-noise ratios (SNR) is set as $20$ dB.
			
			\subsection{Validation of the theoretical results}

			\begin{figure}[!t]
				\centering
				\includegraphics[width=3.5in]{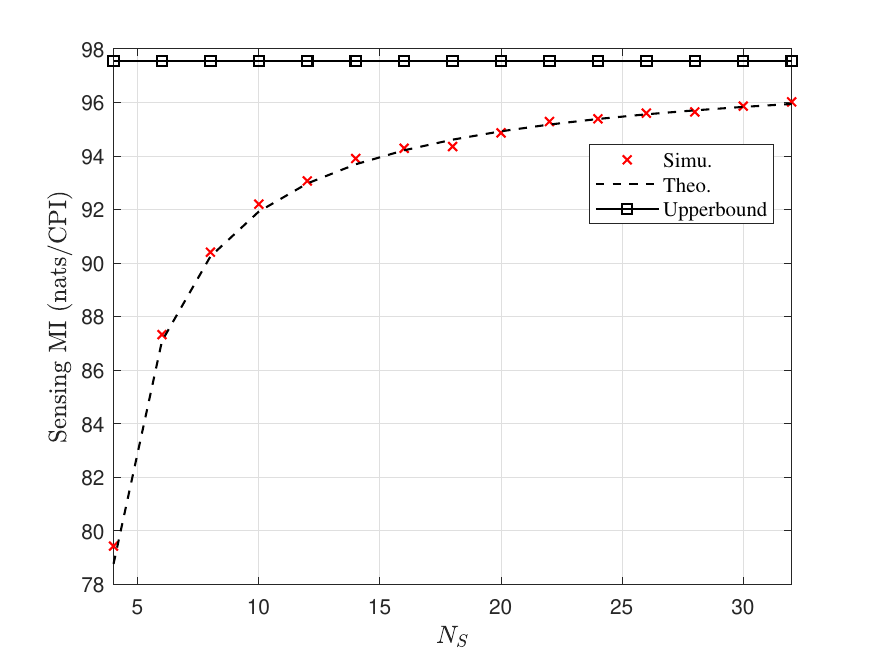}
				\caption{Sensing MI versus the number of frames $N_S$. %(Averaged over 5000 Monte-Carlo trials)
				}
				\label{fig_fitness}
			\end{figure}
			
			\textbf{SMI}: Fig. \ref{fig_fitness} validates the accuracy of the evaluation for SMI in \emph{Proposition \ref{TheoRD}} with $K=7$ targets. The legend `Simu.' denotes the MI obtained by Monte-Carlo simulations, i.e., 
			%$\bar{I}\left(\bm{\eta};\mathbf{y}_s|\mathbf{S}\right)=\frac{1}{N_{mc}}\sum_{i=1}^{N_{mc}}I\left(\bm{\eta};\mathbf{y}_s|\mathbf{S}=\mathbf{S}_{i}\right),$ 
			\begin{equation}\nonumber
				\begin{split}
						\bar{I}\left(\bm{\eta};\mathbf{y}_s|\mathbf{S}\right)=\frac{1}{N_{mc}}\sum_{i=1}^{N_{mc}}I\left(\bm{\eta};\mathbf{y}_s|\mathbf{S}=\mathbf{S}_{i}\right),
					\end{split}
			\end{equation}
			where %$I\left(\mathbf{h}_s;\mathbf{y}_s|\mathbf{S}\right)$ is defined by \eqref{SEMIdefcondS}, 
			$\mathbf{S}_{i}$ denotes the $i$th realization of $\mathbf{S}$ and $N_{mc} = 5000$ represents the number of Monte-Carlo trails. 
			%The number of targets is set as $K=7$. 
			%The number of targets is $K=2$ and the number of receive antennas is $N_R = 4$. Here, we set $N_T$ and $N_S$ increase together with a fixed ratio, i.e., $N_T = 2N_S$. 
			%The number of frames is set as $N_S=8$.
			The legend `Theo.' denotes the theoretical result of the SMI given in \eqref{SEMI0}. 
			The legend `Upperbound' denotes the upper bound in \eqref{ESEMIdefcondS_Jen} and the legend `Lowerbound' represents the lower bound of SMI given in \emph{Proposition \ref{TheoIlb}}. Although \emph{Proposition \ref{TheoRD}} indicates that $I\left(\bm{\eta};\mathbf{y}_s|\mathbf{S}\right)$ will approach $\sum_{j=1}^{K} \bar{\varrho}_{j}(\mathbf{\Phi})$ as $N_S$ goes to infinity, it can be observed that the approximation in \eqref{ESEMIdefcondS_Jen} is very accurate, even when $N_S$ is small. This observation validates the accuracy of \emph{Proposition \ref{TheoRD}}. 
			%This validates \emph{Proposition \ref{TheoRD}}. This observation indicates this approximation
			Meanwhile, a discrepancy consistently exists between SMI and its upper bound, which is more obvious when $N_S$ is small. 
			%As $N_S$ decreases, this discrepancy becomes more obvious.  %This observation is also our motivation to present the tractable expression for the SMI. 
			%

			%\begin{figure}[!t]
			%	\centering
			%	\includegraphics[width=3.5in]{SEMI_fig_0dB_N_T_test1.pdf}
			%	\caption{SEMI versus the number of transmit antennas $N_T$. 
				%	}
			%	\label{fig_fitness_N_T}
			%\end{figure}
			
			%\begin{figure}[!t]
			%	\centering
			%	\includegraphics[width=3.5in]{SEMI_fig_N_T_3_1.pdf}
			%	\caption{ESMI versus the number of transmit antennas $N_T$. 
				%	}
			%	\label{fig_fitness_N_T}
			%\end{figure}
			%
			%Fig. \ref{fig_fitness_N_T} shows the comparison between ‘Simu.’ and ‘Theo.’ results versus the number of transmit antennas $N_T$. We can observe that, 
			%as $N_T$ increases, the discrepancy between the exact value of the average sensing mutual information and its upper bound becomes more pronounced. This phenomenon can be attributed to the fact that the requirement of samples increases with the growth of $N_T$. This observation suggests that the J-ESMI necessitates a significantly large number of samples to accurately represent the actual average performance, potentially leading to high latency. This scenario is further exacerbated by the deployment of extremely large-scale antenna arrays, which is one prominent characteristic of several cutting-edge technologies being considered for the next generation of communication system \cite{9903389}.
			
			\begin{figure}[!t]
				\centering
				\includegraphics[width=3.5in]{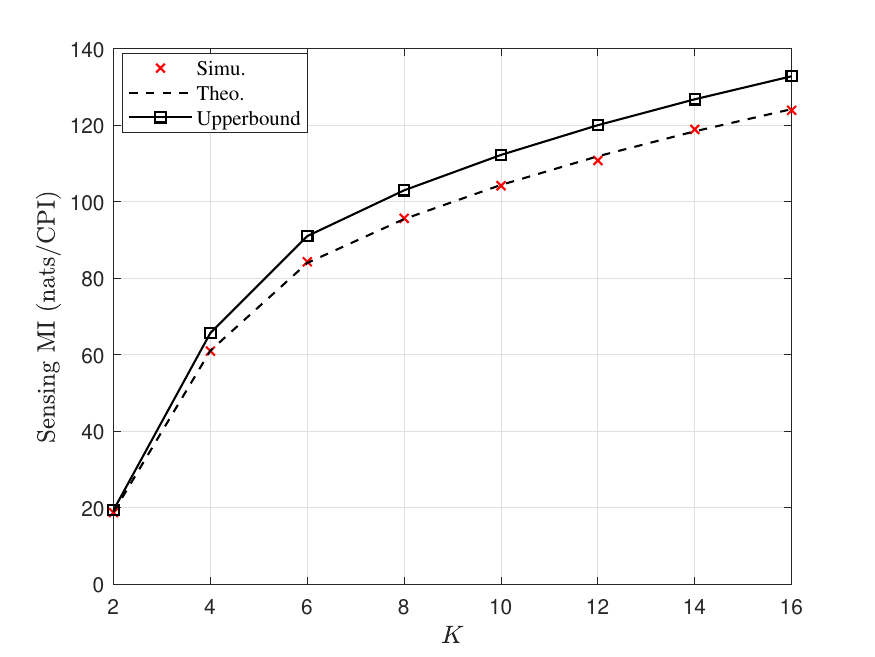}
				\caption{Sensing MI versus the number of targets $K$. 
				}
				\label{fig_fitness_K}
			\end{figure}

			Fig. \ref{fig_fitness_K} compares the theoretical and simulation results with respect to the number of targets $K$, where we set $N_S=8$. 
			We can observe that SMI will increase with $K$. This is because with more targets, there is more information to be estimated. %Each target contributes additional information about the unknown POI, leading to increased uncertainty reduction and higher mutual information.
			However, it is important to note that as $K$ increases, the discrepancy between SMI and its upper bound becomes more pronounced. As discussed in Sec. III. D, this is due to the increase in the sensing DoF loss. Sensing DoF represents the effective number of independent measurements that can be obtained from the observed data. As $K$ increases, more targets introduce more constraints and dependencies among the measurements, resulting in a larger loss of sensing DoF. Therefore, more samples are required to approach the upper bound of SMI, potentially leading to high latency. 
			Moreover, it is important to note that the relationship between $K$ and SMI is not linear. The increase in SMI with $K$ tends to saturate at a larger $K$, because there is a limit on the amount of information that can be extracted from a given number of samples. % from the observed data. %Once $K$ surpasses 6, further improvements in $K$ may not significantly enhance the SMI. 

			\begin{figure}[!t]
				\centering
				\includegraphics[width=3.5in]{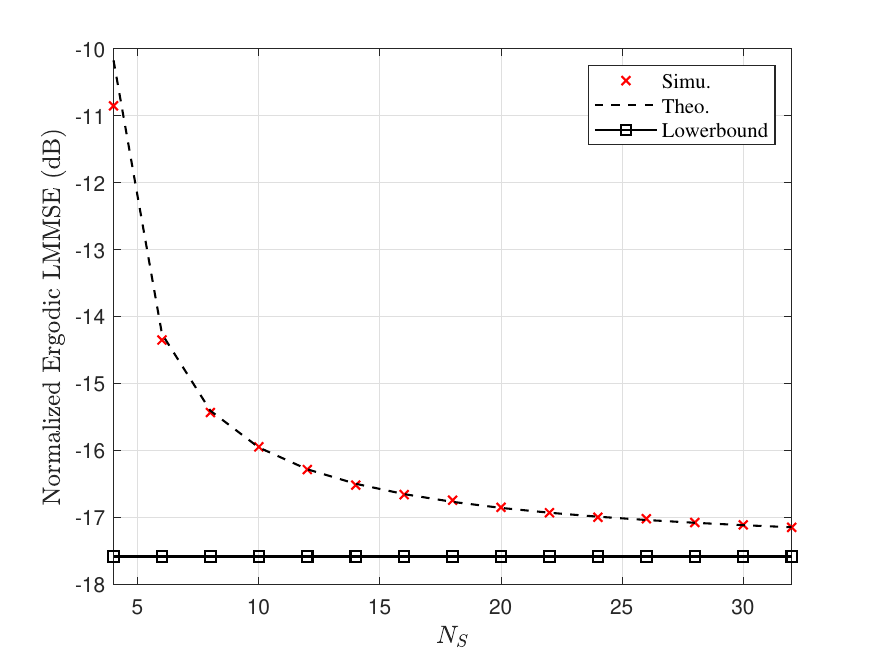}
				\caption{Normalized ELMMSE versus the number of frames $N_S$.}
				\label{fig_fitness2}
			\end{figure}
			\textbf{ELMMSE}: Fig. \ref{fig_fitness2} validates the accuracy of the evaluation for ELMMSE in \emph{Proposition \ref{TheoRD2}} with $K=7$ targets. The normalized ELMMSE is defined as $$\bar{\varepsilon}_{M}=\frac{J_{\mathrm{ELMMSE}}}{\mathbb{E}\;\Vert\mathbf{h}_s\Vert^2}=\frac{J_{\mathrm{ELMMSE}}}{\tr( \mathbf{R})}.$$
			The legend `Lowerbound' denotes the lower bound in \eqref{SEMMSEdef2}. It can be observed that the approximation in \eqref{SEMMSEdefSE} is accurate, particularly when $N_S$ is large, which validates \emph{Proposition \ref{TheoRD2}}.
			With a larger number of samples, the exact value of ELMMSE converges towards its lower bound. 
			%Meanwhile, as $N_S$ increases, the simulation results of ELMMSE may be more accurate compared with the theoretical results of ELMMSE. 
			%This is because the increased number of samples reduces the uncertainty and variability in the SMI, leading to a more accurate approximation of the true value.

			%\begin{figure}[!t]
			%	\centering
			%	\includegraphics[width=3.5in]{SEMI_pro4_fig_test3.pdf}
			%	\caption{Achievable Sensing MI versus the number of frames $N_S$.
				%	}
			%	\label{fig_fitness3}
			%\end{figure}
			
			\begin{figure}[!t]
				\centering
				\includegraphics[width=3.5in]{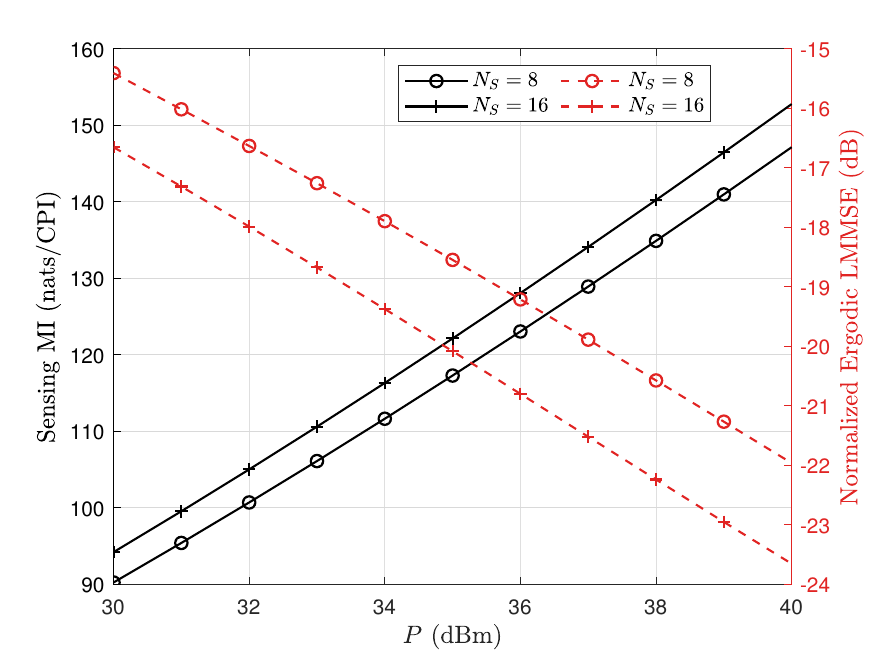}
				\caption{Sensing MI and normalized ELMMSE versus $P$.}
				\label{fig_fitness3}
			\end{figure}
			
			\textbf{Validation of \emph{Proposition \ref{PropCRB_MI_MMSE}}}: Fig. \ref{fig_fitness3} compares SMI and the normalized ELMMSE with different values of $P$. 
			From Fig. \ref{fig_fitness3}, we can observe that as $P$ increases, the SMI increases while the normalized ELMMSE decreases. For example, when $P = 30$ dBm and $N_S = 8$, SMI is about $90.1$ nats/CPI and the normalized ELMMSE is about $-15.3$ dB. When $P = 36$ dBm and $N_S = 8$, SMI is about $123.2$ nats/CPI and the normalized ELMMSE is about $-19.1$ dB. 
			This observation agrees with \emph{Proposition \ref{PropCRB_MI_MMSE}}. 
			
			%Fig. \ref{fig_fitness3} compares $I\left(\mathbf{h}_s;\mathbf{y}_s|\mathbf{S}\right)$ with $\log|\mathbf{R}|- \mathbb{E}_{\mathbf{S}}\log |\mathbf{\Psi}_{\mathrm{BCRB}}|$ under different $N_S$. In particular, $\mathbb{E}_{\mathbf{S}}\log |\mathbf{\Psi}_{\mathrm{BCRB}}|$ is obtained over 5000 Monte-Carlo trials. In \emph{Proposition \ref{PropCRB_MI_MMSE}}, we theoretically prove that $I\left(\mathbf{h}_s;\mathbf{y}_s|\mathbf{S}\right)=\log|\mathbf{R}|- \mathbb{E}_{\mathbf{S}}\log |\mathbf{\Psi}_{\mathrm{BCRB}}|$. From Fig. \ref{fig_fitness3}, we can observe that $I\left(\mathbf{h}_s;\mathbf{y}_s|\mathbf{S}\right)$ can consistently match $\log|\mathbf{R}|- \mathbb{E}_{\mathbf{S}}\log |\mathbf{\Psi}_{\mathrm{BCRB}}|$ for varying values of $N_S$. This observation reveals the physical meaning of SMI, i.e., the maximum information about $\mathbf{h}_s$ at the sensing receiver, which provides a theoretical basis for the precoding design in sensing systems. 
			%the information transfer and improve the performance of sensing systems.

			\subsection{Precoding design for sensing}
			
			\begin{figure}[!t]
				\centering
				\includegraphics[width=3.5in]{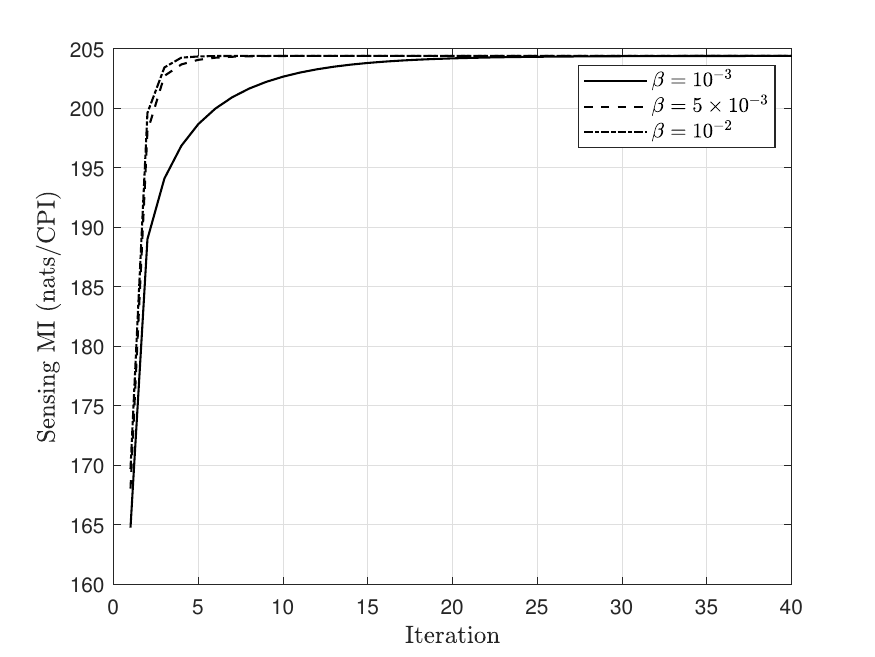}
				\caption{Convergence performance of the proposed GP-based sensing precoding design method.}
				\label{fig_iter}
			\end{figure}
			
			\textbf{Convergence}: 
			%We begin with evaluating the convergence behavior of  \textbf{Algorithm \ref{ALG0}}. 
			Fig. \ref{fig_iter} demonstrates the convergence behavior of the proposed GP-based method, where the number of frames is set as $N_S=8$. 
			We set the step size as $\beta_{m} = 10^{-3}, 5\times  10^{-3}$, and $10^{-2}$, respectively. Note that  the convergence speed can be affected by $\beta_{m}$.
			From Fig. \ref{fig_iter}, we can observe that the proposed GP-based method can converge after several iterations for different values of $\beta_{m}$.
			Therefore, further exploration of different $\beta_{m}$ may be necessary to improve the convergence speed and reach the desired solution. %In particular, to obtain the best convergence speed and ensure convergence to the global optimum, it is often necessary to perform a systematic search over a range of $\beta_{m}$. 
			This search can be done through techniques such as grid search or adaptive methods that dynamically adjust the value of $\beta_{m}$ during the optimization process.
			%Moreover, we can see that the proposed ESMI-oriented optimization can outperform the water-filling optimization based on the J-ESMI. This is because the J-ESMI fails to evaluate the exact value of ESMI when $N_S$ is not large enough. Optimizing J-ESMI can not effectively improve the actual ESMI.

			%\begin{figure*}[!t]
			%	\centering
			%	\subfloat[]{\includegraphics[width=3.5in]{SEMI_comp_N_S_test3_SNR0.pdf}\label{fig_comp_1}}\
			%	\subfloat[]{\includegraphics[width=3.5in]{SEMI_comp_N_S_test3_SNR10.pdf}\label{fig_comp_2}}\
			%	\caption{Sensing MI versus $N_S$. (a) SNR = $0$ dB; (b) SNR = $10$ dB.}
			%	\label{fig_comp_NS}
			%\end{figure*}

			%\begin{figure*}[!t]
			%	\centering
			%	\subfloat[]{\includegraphics[width=3.5in]{SEMI_comp_N_S_test8_P0.pdf}\label{fig_comp_1}}\
			%	\subfloat[]{\includegraphics[width=3.5in]{SEMI_comp_N_S_test8_Pneg10.pdf}\label{fig_comp_2}}\
			%	\caption{Sensing MI versus $N_S$. (a) $P_0$ = $30$ dBm; (b) $P_0$ = $20$ dBm.}
			%	\label{fig_comp_NS}
			%\end{figure*}
			
			\begin{figure}[!t]
				\includegraphics[width=3.5in]{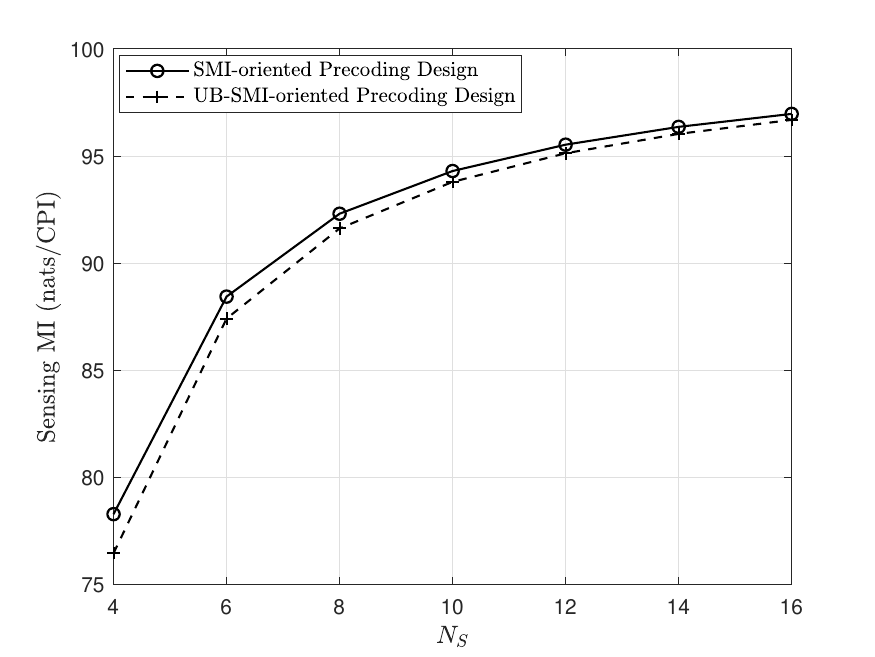}
				\caption{Sensing MI versus $N_S$.}
				\label{fig_comp_NS}
			\end{figure}

			%\begin{figure}[!t]
			%	\centering
			%	\includegraphics[width=3.5in]{SEMI_comp_N_S_test3_2.pdf}
			%	\caption{Sensing MI versus SNR.}
			%	\label{fig_comp}
			%\end{figure}
			%\begin{figure}[!t]
			%	\centering
			%	\includegraphics[width=3.5in]{SEMI_comp_N_S_test3_2.pdf}
			%	\caption{Sensing MI versus SNR.}
			%	\label{fig_comp}
			%\end{figure}

			%\begin{figure}[!t]
			%	\centering
			%	\includegraphics[width=3.5in]{SEMI_comp_SNR_test10.pdf}
			%	\caption{Sensing MI- and Sensing ELMMSE-oriented Precoding Design.}
			%	\label{fig_SMI_SELMMSE}
			%\end{figure}

			%We compare the proposed method with the water-filling method given in \cite{biguesh2006training}, which aimed at optimizing the DSMI. We set $N_S =32$. 
			
			\textbf{Effect of $N_S$}: Note that, as $N_S \to \infty$, SMI will approach its upper bound \eqref{ESEMIdefcondS_Jen}. This allows us to utilize the upper bound as an approximation when $N_S$ is large. 
			The SMI as a function of $N_S$ is shown in Fig. \ref{fig_comp_NS}, where the legend ‘SMI-oriented Precoding Design’ denotes the SMI obtained by the method that maximizes the SMI in \eqref{SEMI0} and the legend ‘UB-SMI-oriented Precoding Design’ represents the SMI obtained by the method that maximizes \eqref{ESEMIdefcondS_Jen}. %The number of targets are set as $K=15$.
			%In Fig. \ref{fig_comp}, 
			%The number of frames and targets are set as $N_S =16$ and $K=15$, respectively. 
			%Figure \ref{fig_comp} illustrates the obtained ESMI using the water-filling method and the proposed method at varying SNR. 
			It can be observed that the proposed method which maximizes the SMI outperforms the one maximizing its upperbound, and the performance gap between the two methods becomes smaller as $N_S$ increases. This can be attributed to the fact that the mismatch between the SMI and its upper bound diminishes as $N_S$ increases. 
			%For sensing, the long-time coherent integration is widely utilized to improve the radar detection ability of a maneuvering target. 
			%However, if the motion states of the targets varies rapidly, CPI should be short and $N_S$ is small. 

\begin{figure}[!t]
				\centering
				\includegraphics[width=3.5in]{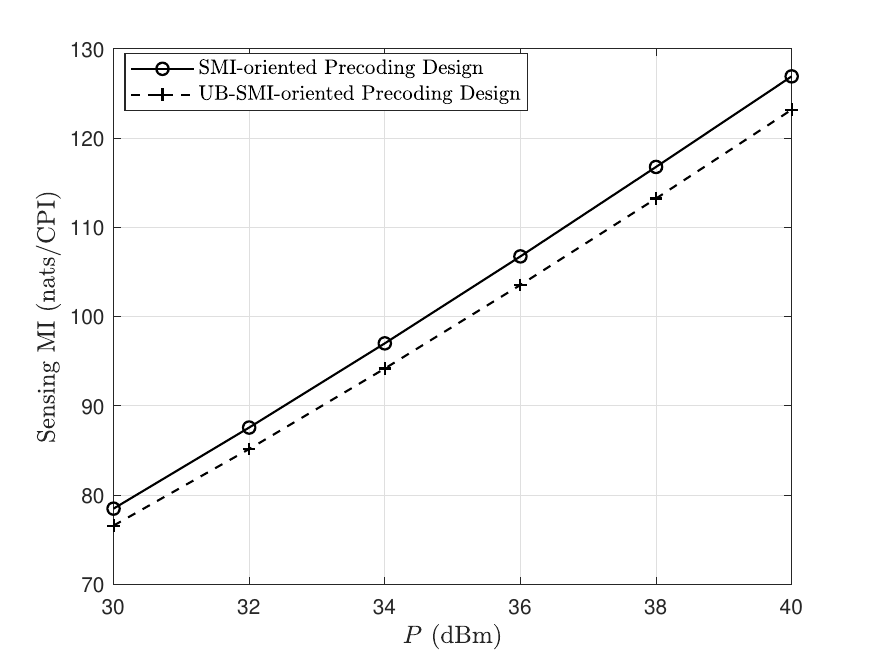}
				\caption{Sensing MI versus $P_0$ ($N_S = 4$).}
				\label{fig_comp_P0}
			\end{figure}

			\textbf{Effect of transmit power $P_0$}: Next, we investigate the impact of transmit power on the optimization results. %In this part, we compared the proposed methods with the existing method. %The legend ‘Water-Filling’ represents the SMI obtained by the method given in \cite{biguesh2006training}, which maximizes \xl{xxx}. 
			It can be observed from Fig. \ref{fig_comp_P0} that SMI will increase with the SNR, achieving a more reliable and accurate estimation of the POIs. This is because SMI represents the amount of information that can be extracted from the observed data regarding POIs. 
			With a higher SNR, the observed data is less corrupted by noise, allowing for more precise estimation of the POIs.
			
			\subsection{Precoding design for ISAC}
			In this section, we will evaluate the performance of the proposed ADMM-based ISAC precoding design method. We consider a communication user equipped with $4$ antennas. %The communication channel is composed of $7$ paths and 
			The distance between the ISAC transmitter and communication receiver is set as $20$ m. 
			The communication requirement is set as $R_0 = 20$ bps/Hz, the number of targets is $K=15$, the transmit power is $P_0=30$ dBm, the penalty parameter is $\rho = 10^4$, and the optimization step size is $\beta_{m} = 1.5 \times 10^{-4}$.
			
\begin{figure}[!t]
				\centering
				\includegraphics[width=3.5in]{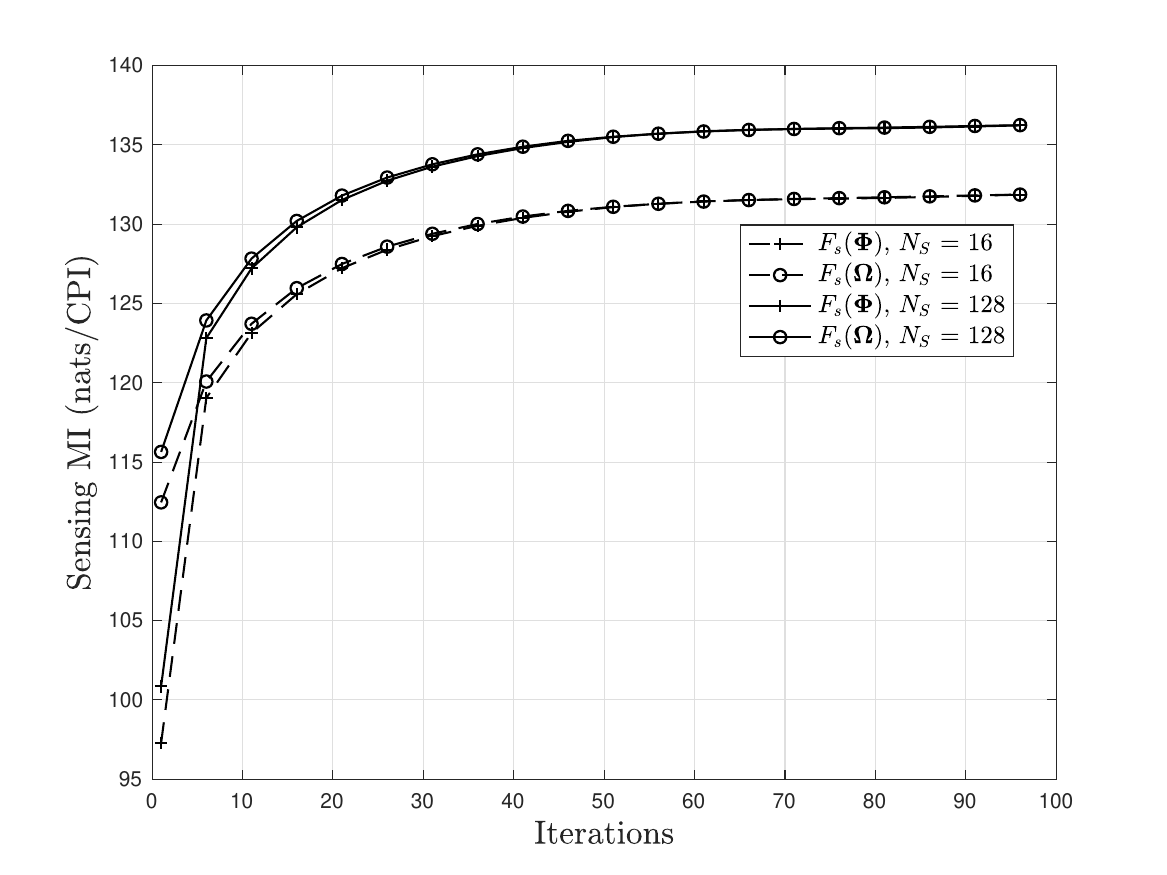}
				\caption{Convergence performance of the proposed ADMM-based ISAC precoding design method.}
				\label{fig_iter_ISAC}
			\end{figure}

			\textbf{Convergence}: We begin with the convergence performance of the proposed ADMM-based ISAC precoding design method. %Note that in the ADMM algorithm, there are 
			Since SMI is a function of $\mathbf{\Phi}$, we denote $\mathcal{F}_s(\mathbf{\Phi})=I\left(\bm{\eta};\mathbf{y}_s|\mathbf{S}\right)$. 
			Fig. \ref{fig_iter_ISAC} shows the SMI with respect to the dual variables  $\mathbf{\Phi}$ and $\mathbf{\Omega}$ over the ADMM iterations, i.e., $\mathcal{F}_s(\mathbf{\Phi})$ and $\mathcal{F}_s(\mathbf{\Omega})$, respectively. We can observe from Fig. \ref{fig_iter_ISAC} that the proposed algorithm can converge within about 50 iterations.

			\begin{figure}[!t]
				\centering
				\includegraphics[width=3.5in]{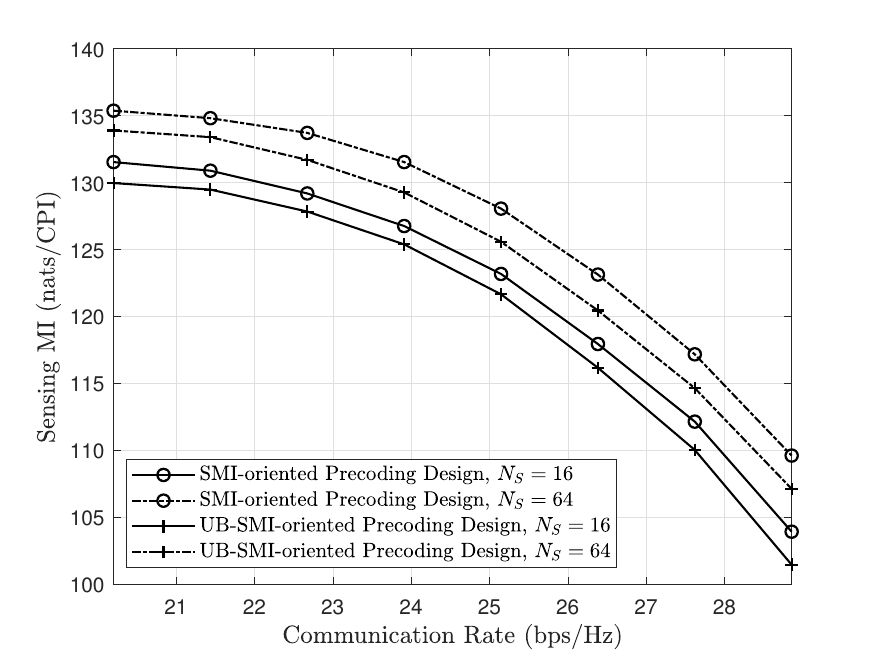}
				\caption{Sensing and Communication tradeoff under different precoding schemes.}
				\label{fig_tradeoff}
			\end{figure}

			\textbf{Tradeoff between sensing and communication}: 
			Next, we will show the performance tradeoff between sensing and communication in Fig.  \ref{fig_tradeoff}. To this end, MI is utilized as a unified metric for evaluating both sensing and communication performance. Specifically, the communication MI is used as a performance metric for communication, while SMI serves as a performance metric for sensing. 
			%The unit of communication rate ``bps/Hz'' denotes ``bits per second per Hertz".  %which is widely used to measure the spectral efficiency of a communication system.
			% with different sensing metrics. 
			From Fig. \ref{fig_tradeoff}, we can observe that as the communication rate increases, the SMI will decrease. 
			This is because there is an inherent competition between sensing and communication in an ISAC system. 
			%On the one hand, sensing requires resources to probe the target for extracting useful information about the sensing parameters. On the other hand, communication also requires resources for transmitting data to users. 
			Additionally, the results obtained by optimizing the SMI surpass those obtained by optimizing its upper bound in \eqref{ESEMIdefcondS_Jen}. %For example, when the communication rate is 41.9 bps/Hz, the SMI of `SMI-oriented precoding design' is 6.7 nats/CPI higher than that of `UB-SMI-oriented precoding design'. 

			%\begin{figure}[!t]
			%	\centering
			%	\includegraphics[width=3.5in]{SEMI_tradeoff_NS_test2.pdf}
			%	\caption{Sensing and Communication tradeoff versus $N_S$.}
			%	\label{fig_tradeoff_L}
			%\end{figure}

			%\clearpage
			\section{Conclusion}
			This paper contributed to the unification of sensing and communication performance metrics. For that purpose, we started by deriving an explicit expression for SMI with random signals using random matrix theory. Based on these results, we established the connections between SMI, EMMSE, ELMMSE, and EBCRB. These connections demonstrated that SMI can serve as a bridge to connect communication and sensing metrics, providing a unified framework for the analysis and design of ISAC systems. 
			Finally, two optimization-based methods were proposed to maximize the SMI by designing the precoder for both sensing-only and ISAC scenarios. 
			Simulation results validated the accuracy of the theoretical results and the effectiveness of the proposed optimization-based methods. 
			
			\appendices
			\section{Proof of Proposition \ref{TheoRD}}\label{proofTheoRD}
			This proof is equivalent to analyze the asymptotic behavior of the following random variable
			\begin{equation}\label{SEMI00}
				\begin{split} 
					\varrho =\log \left\vert\mathbf{I}+ \sigma_s^{-2}\mathbf{R}_R \otimes (\mathbf{R}_{T}^{\frac{1}{2}}\mathbf{F}\mathbf{S}\mathbf{S}^\HH \mathbf{F}^\HH\mathbf{R}_{T}^{\frac{1}{2}} ) \right\vert.
				\end{split}
			\end{equation}
			In fact, \emph{Proposition} \ref{TheoRD} is an extension of 
			\cite[Theorems 1]{hachem2008new}, which is summarized in the following lemma. 
			
			\begin{lemma}[{\cite[Theorems 1]{hachem2008new}}]\label{lemmatheo1}
				Given two deterministic diagonal matrices $\mathbf{D} \in \mathcal{R}^{N\times N}$ and $\widetilde{\mathbf{D}} \in \mathcal{R}^{n\times n}$, we can define $\mathbf{Z}=\mathbf{D}^{\frac{1}{2}} \widetilde{\mathbf{Z}} \widetilde{\mathbf{D}}^{\frac{1}{2}}$, where $\widetilde{\mathbf{Z}}$ has i.i.d. entries following complex Gaussian distribution $\mathcal{CN}(0,\frac{1}{n})$. Then, we have 
				\begin{equation}\label{oldlemma}
					\begin{split}
						\mathbb{E}  \log \left\vert\rho \mathbf{Z}\mathbf{Z}^\HH+\mathbf{I}\right\vert = V_{n}(\rho) + \mathcal{O}\left(\frac{1}{n}\right),
					\end{split}
				\end{equation}
				where  
				\begin{equation}\nonumber
					\begin{split}
						V_{n}(\rho) = & \log \left\vert\mathbf{I}+\rho \delta(\rho) \widetilde{\mathbf{D}}\right\vert + \log \left\vert\mathbf{I}+\rho \tilde{\delta}(\rho) \mathbf{D}\right\vert -n \rho \delta(\rho)\tilde{\delta}(\rho),
					\end{split}
				\end{equation}
				and  $(\delta(\rho),\tilde{\delta}(\rho))$ is the unique positive solution of the following fixed-point equations
				\begin{equation}\label{oldfp}
					\begin{split}
						\left\{
						\begin{matrix}
							\delta = \frac{1}{n} \tr  \left(\mathbf{D}\left(\mathbf{I}+\rho \tilde{\delta} \mathbf{D}\right)^{-1}\right)\\
							\tilde{\delta} = \frac{1}{n} \tr \left(\widetilde{\mathbf{D}}\left(\mathbf{I}+\rho \delta  \widetilde{\mathbf{D}}\right)^{-1}\right)
						\end{matrix}\right. .
					\end{split}
				\end{equation}	
			\end{lemma}

			The main idea of this proof is to recast $I\left(\mathbf{h}_s;\mathbf{y}_s|\mathbf{S}\right)$ into an extended model which fits into the framework of \cite{hachem2008new}.
			
			First, by utilizing the property of block diagonal matrix and the equality  $\det(\mathbf{I}+\mathbf{A}\mathbf{B})=\det(\mathbf{I}+\mathbf{B}\mathbf{A})$, (\ref{SEMI00}) can be rewritten as 
			%$\varrho =\sum_{j=1}^{K} \varrho_j$, 
			\begin{equation}\label{rhojsum}
				\varrho =\sum_{j=1}^{K} \varrho_j,
			\end{equation}
			where $$\varrho_j=\log \det\left(\mathbf{I}_{N_T} + \lambda_{R,j} \mathbf{R}_{T}^{\frac{1}{2}}\mathbf{F}\mathbf{S}\mathbf{S}^\HH \mathbf{F}^\HH\mathbf{R}_{T}^{\frac{1}{2}} \right).$$
			%By performing the eigen-value decomposition on $\mathbf{D}$, and $\mathbf{T}$, we have $\mathbf{D}=\mathbf{U}_D \widehat{\mathbf{D}} \mathbf{U}_D^\HH$ and $\mathbf{T}=\mathbf{U}_T \widehat{\mathbf{T}} \mathbf{U}_T^\HH$, respectively. Note that $\widehat{\mathbf{D}}$ and $\widehat{\mathbf{T}}$ are diagonal and $\mathbf{U}_D$ and $\mathbf{U}_T$ are unitary. 
			By performing the eigenvalue decomposition on $ \mathbf{R}_{T}^{\frac{1}{2}}\mathbf{F}$, we have $\mathbf{R}_{T}^{\frac{1}{2}}\mathbf{F}=\mathbf{U}_T \widetilde{\mathbf{\Lambda}}_{T,0} \mathbf{V}_T^\HH$, where $\mathbf{U}_T \in \mathbb{C}^{N_T\times N_T}$ and $\mathbf{V}_T \in \mathbb{C}^{K\times K}$ denote the left- and right-singular vectors, respectively. Meanwhile, $\widetilde{\mathbf{\Lambda}}_T$ is a rectangular diagonal matrix with non-negative real numbers on the diagonal, i.e.,
			\begin{equation}
				\begin{split}
					\widetilde{\mathbf{\Lambda}}_{T,0}=\left[\begin{matrix}
						\mathbf{\Lambda}_T^{\frac{1}{2}} \\
						\mathbf{0} 
					\end{matrix}
					\right],
				\end{split}
			\end{equation}
			with $\mathbf{\Lambda}_T \in \mathbb{C}^{K\times K}$ denotes the eigenvalue matrix of $\mathbf{T}\triangleq \mathbf{R}_{T}^{\frac{1}{2}}\mathbf{F}\mathbf{F}^\HH\mathbf{R}_{T}^{\frac{1}{2}}$, i.e., $\mathbf{T}=\mathbf{U}_T \widetilde{\mathbf{\Lambda}}_T \mathbf{U}_T^\HH$,
			where
			\begin{equation}
				\begin{split}
					\widetilde{\mathbf{\Lambda}}_T=\left[\begin{matrix}
						\mathbf{\Lambda}_T & \mathbf{0}\\
						\mathbf{0} & \mathbf{0}
					\end{matrix}
					\right].
				\end{split}
			\end{equation}
			
			Note that, with any unitary matrices $\mathbf{U}$ and $\mathbf{V}$, the random matrices $\mathbf{U} \mathbf{S} \mathbf{V}$ and $\mathbf{S}$ are statistically equivalent \cite{hachem2008new}. Thus, by omitting some constants independent of $\mathbf{S}$, the asymptotic behavior of $\varrho_j$ is equivalent to that of the RV %$\hat{\varrho}_j = \log \det\left(\mathbf{I}_{K} + \lambda_{R,j} \mathbf{\Lambda}_{T}^{\frac{1}{2}}\mathbf{S}\mathbf{S}^\HH \mathbf{\Lambda}_{T}^{\frac{1}{2}} \right)$.
			\begin{equation}\label{varrhol0}
				\hat{\varrho}_j = \log \det\left(\mathbf{I}_{K} + \lambda_{R,j} \mathbf{\Lambda}_{T}^{\frac{1}{2}}\mathbf{S}\mathbf{S}^\HH \mathbf{\Lambda}_{T}^{\frac{1}{2}} \right).
			\end{equation}
			By invoking $\mathbf{D}=\mathbf{\Lambda}_{T}$, $\widetilde{\mathbf{D}}=\mathbf{I}_{N_S}$, $n=N_S$, and $\rho = \lambda_{R,j}$ into \eqref{oldfp}, we have $\tilde{\delta}(\rho)=1/(1+\rho \delta )$ such that the fixed-point equations in \eqref{oldfp} reduce to a linear one with respect to $\delta$, i.e.,
			\begin{equation}\label{deltadef1}
				\delta(\rho) = \frac{1}{N_S} \tr \left(\mathbf{\Lambda_{T}}\left(\mathbf{I}_{K}+\frac{\rho}{1+\rho\delta(\rho)} \mathbf{\Lambda_{T}}\right)^{-1}\right).
			\end{equation}
			Furthermore, from \emph{Lemma 1}, we can obtain that, as $N_S,N_T \to \infty$, one has
			\begin{equation}\label{exprhoj}
				\begin{split}
					\mathbb{E}(\hat{\varrho}_j) =\bar{\varrho}_j+\mathcal{O}\left(\frac{1}{N_S}\right), j =1,2,\cdots,
				\end{split}
			\end{equation}
			where 
			\begin{equation}\label{averrhoj1}
				\begin{split}
					&\bar{\varrho}_j = \log \left\vert\mathbf{I}_{K}+\frac{\lambda_{R,j}}{1+\lambda_{R,j}\delta(\lambda_{R,j})} \mathbf{\Lambda_{T}} \right\vert\\
					&+N_S \log (1+\lambda_{R,j}\delta(\lambda_{R,j}))-\frac{N_S\lambda_{R,j}\delta(\lambda_{R,j})}{1+\lambda_{R,j}\delta(\lambda_{R,j})}.
				\end{split}
			\end{equation}
			
			Then, by utilizing the property of the unitary matrix $\mathbf{U}_T$, \eqref{averrhoj1} and \eqref{deltadef1} are equivalent to 
			\eqref{averrhoj0} and \eqref{fixeq1}, respectively, which completes the proof.

			\section{Proof of Lemma \ref{lemmaderiveRs}}
			\label{prooflemmaderiveRs}
			%To simplify the notation, we denote $\delta(\lambda_{R,j})$ as $\delta$ in this proof. 
			From \eqref{fixeq1}, we have $\delta(\rho) = \frac{1}{N_S} \tr (\mathbf{T}\mathbf{M}_T),$ 
			%	\begin{equation}
				%		\begin{split}
					%		\delta(\rho) = \frac{1}{N_S} \tr \mathbf{T}\mathbf{M}_T,
					%	\end{split}
				%	\end{equation}
			where  $\mathbf{M}_T(\rho)\triangleq\left(\mathbf{I}_{N_S}+\alpha(\rho) \mathbf{T}\right)^{-1}$ and $\alpha(\rho)=\frac{\rho}{1+\rho\delta(\rho)}$. 
			%\begin{equation}
			%	\begin{split}
				%\mathbf{M}_T(\rho)\triangleq\left(\mathbf{I}_{N_S}+\alpha(\rho) \mathbf{T}\right)^{-1},
				%	\end{split}
			%\end{equation}
			%and 
			%\begin{equation}
			%	\begin{split}
				%		\alpha(\rho)=\frac{\rho}{1+\rho\delta(\rho)}.
				%	\end{split}
			%\end{equation}
			The derivative of $\alpha(\rho)$ with respect to the $(m,n)$th entry of $\mathbf{F}$, denoted by $F_{m,n}^*$, is given by
			\begin{equation}\label{derproof001}
				\begin{split}
					\alpha_{m,n}'(\rho)&=\frac{\partial \alpha(\rho)}{\partial \Phi_{m,n}^*}=-\alpha^2(\rho) [\mathbf{\Delta}_{\rho}'(\mathbf{\Phi})]_{m,n}.
				\end{split}
			\end{equation}
			%Note that $\mathbf{T}$ is dependent on $\mathbf{F}$. 
			The derivative of $\mathbf{T}$ with respect to $F_{m,n}^*$ can be expressed as
			\begin{equation}\label{derproof002}
				\begin{split}
					[{\mathbf{T}'}]_{m,n}=\frac{\partial \mathbf{T}}{\partial \Phi_{m,n}^*}= \mathbf{R}_T^{\frac{1}{2}} \mathbf{e}_n \mathbf{e}_m^\TT \mathbf{R}_T^{\frac{1}{2}}.
				\end{split}
			\end{equation}
			Therefore, we have
			\begin{equation}\nonumber
				\begin{split}
					&[\mathbf{\Delta}_{\rho}'(\mathbf{\Phi})]_{m,n}=\frac{\partial \frac{1}{N_S}\tr \left(\mathbf{T}\left(\mathbf{I}_{N_S}+\alpha(\rho)\mathbf{T}\right)^{-1}\right)}{\partial \Phi_{m,n}^*}\\
					&=\frac{1}{N_S}\tr \left([{\mathbf{T}'}]_{m,n} \mathbf{M}_{T}(\rho)\right) \\
					&\quad - \frac{1}{N_S}\tr \left(\mathbf{T} \mathbf{M}_{T}(\rho)
					\left(\alpha_{m,n}'(\rho) \mathbf{T} + \alpha(\rho) [{\mathbf{T}'}]_{m,n} \right)\mathbf{M}_{T}(\rho)\right)\\
					%&= \frac{\alpha^2(\rho)\tr\left(\mathbf{T} \mathbf{M}_{T}(\rho)\right)^2 }{N_S} [\mathbf{\Delta}_{\rho}'(\mathbf{\Phi})]_{m,n}\\
					%&\quad +\frac{1}{N_S}\tr \left[ \left(\mathbf{I}-\alpha(\rho)\mathbf{T}\mathbf{M}_T(\rho)\right)[{\mathbf{T}'}]_{m,n} \mathbf{M}_{T}(\rho)\right]\\
					&=  \frac{\alpha^2(\rho)\tr\left(\mathbf{T} \mathbf{M}_{T}(\rho)\right)^2 }{N_S} [\mathbf{\Delta}_{\rho}'(\mathbf{\Phi})]_{m,n}+\frac{[\mathbf{R}_T^{\frac{1}{2}}\mathbf{M}_T^2(\rho)\mathbf{R}_T^{\frac{1}{2}}]_{m,n}}{N_S}.
				\end{split}
			\end{equation}
			The solution to the above linear equation is given by 
			\begin{equation}\label{derivaLER2}
				\begin{split}
					[\mathbf{\Delta}_{\rho}'(\mathbf{\Phi})]_{m,n} = \frac{[\mathbf{R}_T^{\frac{1}{2}}\mathbf{M}_T^2(\rho)\mathbf{R}_T^{\frac{1}{2}}]_{m,n}}{N_S-\alpha^2(\rho)\tr\left(\mathbf{T}(\mathbf{\Phi}) \mathbf{M}_{T}(\rho)\right)^2 }.
				\end{split}
			\end{equation}
			%			where $\mathbf{M}_{T}(\rho) =\left(\mathbf{I}+\frac{\rho}{1+\rho\delta(\rho)} \mathbf{T}(\mathbf{\Phi})\right)^{-1}$.
			%			\begin{equation}\label{derivaLER}
				%				\begin{split}
					%						&[\mathbf{\Delta}_{\rho}'(\mathbf{\Phi})]_{m,n} \\
					%						&= \frac{\left[\mathbf{R}_T^{\frac{1}{2}}\left(\mathbf{I}+\frac{\rho}{1+\rho\delta(\rho)} \mathbf{T}(\mathbf{\Phi})\right)^{-2}\mathbf{R}_T^{\frac{1}{2}}\mathbf{F}\right]_{m,n}}{N_S-\alpha^2(\rho)\tr\left(\mathbf{T}(\mathbf{\Phi}) \left(\mathbf{I}+\frac{\rho}{1+\rho\delta(\rho)} \mathbf{T}(\mathbf{\Phi})\right)^{-1}\right)^2 }.
					%					\end{split}
				%			\end{equation}
			By consolidating all $[\mathbf{\Delta}_{\rho}'(\mathbf{\Phi})]_{m,n}$ into one matrix, we can obtain \eqref{derivaLER}.

			\section{Proof of Proposition \ref{TheoIlb}}
			\label{proofTheoIlb}
			
			Since $\log (1+x)-\frac{x}{1+x} \geq 0$ for $x\geq0$, we have
			\begin{equation}\label{varrhojAB}
				\begin{split}	
					&\bar{\varrho}_{j}(\mathbf{\Phi})\geq\log \left\vert\mathbf{I}_{N_T}+\frac{\lambda_{R,j}}{1+\lambda_{R,j}\delta(\lambda_{R,j})} \mathbf{T}(\mathbf{\Phi}) \right\vert \\
					& \overset{(a)}{\geq} \frac{1}{1+\lambda_{R,j}\delta(\lambda_{R,j})}\log \left\vert\mathbf{I}_{N_T}+\lambda_{R,j} \mathbf{T}(\mathbf{\Phi}) \right\vert,
				\end{split}	
			\end{equation}
			where step (a) follows the Jensen's inequality. Recalling \eqref{fixeq1}, we have
			\begin{equation}
				\delta(\rho) = \frac{1}{N_S}\sum_{i=1}^{r_T}   \lambda_{T,i}\left(1+\frac{\rho}{1+\rho\delta(\rho)} \lambda_{T,i}\right)^{-1},
			\end{equation}
			where $r_T = \mathrm{rank}(\mathbf{T}) \leq \min\{K,\mathrm{rank}(\mathbf{\Phi})\}$ and $\lambda_{T,i}$ denotes the $i$th non-zero eigenvalue of $\mathbf{T}$. Given $a>0$,  $\frac{x}{1+ax}$ is monotonically increasing for $x\geq0$. Thus, we have
			\begin{equation}
				\delta(\rho) \leq \lim_{\lambda_{T,i} \to \infty}\delta(\rho)= \frac{r_T(1+\rho\delta(\rho))}{\rho N_S} .
			\end{equation}
			Given  $N_S \geq K \geq r_T$, we have 
			\begin{equation}\label{deltaub}
				\delta(\rho) \leq \frac{r_T}{\rho(N_S-r_T)}. %\frac{1}{\rho}\cdot\frac{r_T}{N_S-r_T}.
			\end{equation}
			By substituting \eqref{deltaub} into \eqref{varrhojAB}, we have
			\begin{equation}\label{varrhojAB2}
				\begin{split}	
					&\bar{\varrho}_{j}(\mathbf{\Phi})\geq \frac{N_S -r_T}{N_S}\log \left\vert\mathbf{I}_{N_T}+\lambda_{R,j} \mathbf{T}(\mathbf{\Phi}) \right\vert.
				\end{split}	
			\end{equation}
			By taking the summation of \eqref{varrhojAB2} over index $j$, \eqref{Ilower bound} can obtained. 
			
			\section{Proof of Proposition \ref{TheoRD2}}\label{proofTheoRD2}
			This proof is equivalent to analyze the asymptotic behavior of the following random variable
			\begin{equation}\label{SELMMSE00}
				\begin{split} 
					\varkappa =\tr \left[ \mathbf{R}\left(\mathbf{I}+ \sigma_s^{-2}\left(\mathbf{R}_R \otimes \mathbf{R}_T^{\frac{1}{2}}\mathbf{F}\mathbf{S}\mathbf{S}^\HH\mathbf{F}^\HH\mathbf{R}_T^{\frac{1}{2}}\right)\right)^{-1}\right].
				\end{split}
			\end{equation}
			In fact, \emph{Proposition} \ref{TheoRD2} is an extension of \cite[Proposition 5]{hachem2008new} and  \cite[Theorems 3]{hachem2008new}, which are summarized in the following lemmas.  
			\begin{lemma}[{\cite[Proposition 5]{hachem2008new}}]\label{lemmaprop5}
				Given two diagonal matrices $\mathbf{D} \in \mathcal{R}^{N\times N}$ and $\widetilde{\mathbf{D}} \in \mathcal{R}^{n\times n}$, we can define $\mathbf{Z}=\mathbf{D}^{\frac{1}{2}} \widetilde{\mathbf{Z}} \widetilde{\mathbf{D}}^{\frac{1}{2}}$, where $\widetilde{\mathbf{Z}}$ has i.i.d. entries with distribution $\mathcal{CN}(0,\frac{1}{n})$.  Further, for all $t>0$, let $\mathbf{A}$ be a uniformly bounded diagonal $N \times N$ matrix, we have
				\begin{equation}\label{oldpro5}
					\mathbb{E} \; \tr \left(\mathbf{A} \left(\mathbf{I} + \rho \mathbf{Z}\mathbf{Z}^\HH\right)^{-1}\right) = \tr \left(\mathbf{A} \mathbf{Q}\right) + \mathcal{O}\left(\frac{1}{N_S}\right),
				\end{equation}		
				where	 
				\begin{equation}
					\begin{split}
						&\widetilde{\beta} = \frac{1}{n} \tr \left(\widetilde{\mathbf{D}} \left(\mathbf{I} + \rho \beta\widetilde{\mathbf{D}} \right)^{-1}\right),\\
						&\mathbf{Q} = \left(\mathbf{I}+\widetilde{\beta} \mathbf{D}\right)^{-1},\\
						&\beta = \frac{1}{n} \tr \left(\mathbf{D} \mathbb{E}\left(\mathbf{I} + \rho \mathbf{Z}\mathbf{Z}^\HH\right)^{-1}\right),\\
						&\widetilde{\mathbf{Q}} = \left(\mathbf{I}+\beta \widetilde{\mathbf{D}}\right)^{-1}.\\
					\end{split}
				\end{equation}	  
			\end{lemma}

			\begin{lemma}[{\cite[Theorem 3]{hachem2008new}}]\label{lemmatheo3}
				With the same setting of \emph{Lemma \ref{lemmatheo1}}, we have
				\begin{equation}\label{oldtheo3}
					\frac{1}{n} \mathbf{A} \mathbf{Q} =  	\frac{1}{n} \mathbf{A} \mathbf{M}+ \mathcal{O}\left(\frac{1}{n^2}\right),
				\end{equation}
				where
				\begin{equation}\label{oldtheo300}
					\mathbf{M} = \left(\mathbf{I}+\rho \tilde{\delta}(\rho) \mathbf{D}\right)^{-1}.
				\end{equation}
				
			\end{lemma}
			
			The main idea of this proof is to recast $J_{G}$ into an extended model which fits into the framework of \cite{hachem2008new}. 
			
			First, $\varkappa$ can be reformulated by 
			\begin{equation}\label{SELMMSE00}
				\begin{split} 
					\varkappa = \sum_j^{K} \varkappa_j,
				\end{split}
			\end{equation}
			where $\varkappa_j= \sigma_s^2\lambda_{R,j} \tr( \mathbf{R}_{T} (\mathbf{I}_{N_T} + \lambda_{R,j} \mathbf{R}_{T}^{\frac{1}{2}}\mathbf{F}\mathbf{S}\mathbf{S}^\HH \mathbf{F}^\HH\mathbf{R}_{T}^{\frac{1}{2}} )^{-1}).$

			By omitting some constants independent of $\mathbf{S}$, the asymptotic behavior of $\varrho_j$ is equivalent to that of the RV 
			\begin{equation}
				\begin{split}
					\hat{\varkappa}_j = \sigma_s^2\lambda_{R,j}  \tr \left( \mathbf{A}\left(\mathbf{I}_{K} + \lambda_{R,j} \mathbf{\Lambda}_{T}^{\frac{1}{2}}\mathbf{S}\mathbf{S}^\HH \mathbf{\Lambda}_{T}^{\frac{1}{2}} \right)^{-1}\right), 
				\end{split}
			\end{equation}
			where $\mathbf{A} =\mathbf{U}_T^\HH \mathbf{R}_{T} \mathbf{U}_T$. %According to \cite[Proposition 5]{hachem2008new}, we have 
			By invoking $\mathbf{D}=\mathbf{\Lambda}_{T}$, $\widetilde{\mathbf{D}}=\mathbf{I}_{N_S}$, $n=N_S$, and $\rho = \lambda_{R,j}$ into \emph{Lemma \ref{lemmaprop5}}, we have
			\begin{equation}\label{appr0001}
				\begin{split}
					\mathbb{E} (\hat{\varkappa}_j) &= \sigma_s^2\lambda_{R,j} \mathbb{E} \;\tr \left( \mathbf{A}\left(\mathbf{I}_{K} + \lambda_{R,j} \mathbf{\Lambda}_{T}^{\frac{1}{2}}\mathbf{S}\mathbf{S}^\HH \mathbf{\Lambda}_{T}^{\frac{1}{2}} \right)^{-1}\right)\\
					&=\sigma_s^2\lambda_{R,j} \tr  \left(\mathbf{A}\mathbf{Q}_j\right) +\mathcal{O}\left(\frac{1}{N_S}\right),
				\end{split}
			\end{equation}
			where 
			\begin{equation}
				\begin{split}
					\mathbf{Q}_j = \left( \mathbf{I}_{K}+\frac{\lambda_{R,j}}{1+\lambda_{R,j}\beta_j} \mathbf{\Lambda_{T}}  \right)^{-1},
				\end{split}
			\end{equation}
			and
			\begin{equation}
				\begin{split}
					\beta_j = \frac{1}{N_S}\tr\left[\mathbf{\Lambda}_T \left(\mathbb{E}\left(\mathbf{I}_{N_T} + \lambda_{R,j} \mathbf{\Lambda}_T^{\frac{1}{2}}\mathbf{S}\mathbf{S}^\HH \mathbf{\Lambda}_T^{\frac{1}{2}} \right)^{-1}\right)\right].
				\end{split}
			\end{equation}
			
			\emph{Lemma \ref{lemmatheo3}} together with \eqref{appr0001} implies that
			\begin{equation}\label{appr0002}
				\begin{split}\nonumber
					& \tr \left( \mathbf{A}\mathbf{Q}_j\right) \\
					& =  \tr\left(\mathbf{A} \left(\mathbf{I}_{N_T}+\frac{\lambda_{R,j}}{1+\lambda_{R,j}\delta(\lambda_{R,j})} \mathbf{\Lambda}_T\right)^{-1}\right)+\mathcal{O}\left(\frac{1}{N_S}\right)\\
					& = \tr\;  \mathbf{R}_T \mathbf{M}_{T,j}+\mathcal{O}\left(\frac{1}{N_S}\right).
				\end{split}
			\end{equation}

			%	\section{Proof of Proposition \ref{nongaussianbridge}}
			%	\label{nongaussianbridgeproof}
			%	Define 
			%	$	\mathbf{\Psi}_I = \frac{\partial I(\mathbf{h}_s;\mathbf{y}_s|\mathbf{S})}{\partial \sigma_s^{-2}\left(\mathbf{I}_{N_R} \otimes \mathbf{X}\right)\left(\mathbf{I}_{N_R} \otimes \mathbf{X}\right)^\HH}$ for any realization of $\mathbf{S}$. 
			%	According to \cite[Theorem 2]{reeves2018mutual}, we have
			%	$\mathbf{\Psi}_{\mathrm{BCRB}}  \preceq \mathbf{\Psi}_I  \preceq \mathbf{\Psi}_{\mathrm{LMMSE}}$.
			%	By taking the expectation over $\mathbf{S}$ on $\mathbf{\Psi}_{\mathrm{BCRB}}$, $\mathbf{\Psi}_I$, and $\mathbf{\Psi}_{\mathrm{LMMSE}}$, \eqref{nongaussianineq} can be obtained.
			
			\section{Proof of Proposition \ref{PropEQ}}
			\label{proofPropEQ}

			Based on \emph{Lemma \ref{lemmaderiveRs}},  $\nabla_{\mathbf{\Phi}}\bar{\varrho}_{j}(\mathbf{\Phi})$ can be obtained by%\footnote{Note that we simplify the notation by omitting the argument of $\mathbf{M}_{T,j}(\mathbf{\Phi})$, $\alpha_j(\mathbf{\Phi})$, $\mathbf{T}(\mathbf{\Phi})$, and $\mathbf{\Delta}_{\lambda_{R,j}}'(\mathbf{\Phi})$ in \eqref{varrhograd}.}
			\begin{align}
				&\nabla_{\mathbf{\Phi}}\bar{\varrho}_{j}(\mathbf{\Phi})=-\alpha_j^2\tr\left(\mathbf{M}_{T,j}\mathbf{T}\right)\mathbf{\Delta}_{\lambda_{R,j}}'+ \alpha_j \mathbf{R}_T^{\frac{1}{2}}\mathbf{M}_{T,j}\mathbf{R}_T^{\frac{1}{2}}  \notag \\
				&+N_S\alpha_j \mathbf{\Delta}_{\lambda_{R,j}}'-\frac{N_S \lambda_{R,j}}{\left(1+\lambda_{R,j}\delta(\lambda_{R,j})\right)^2} \mathbf{\Delta}_{\lambda_{R,j}}', \label{varrhograd}
			\end{align}
			where $\alpha_j = \alpha(\lambda_{R,j})$ and 
			%	$\mathbf{M}_{T,j}(\mathbf{\Phi})=\left(\mathbf{I}_{N_S}+\alpha_j(\mathbf{\Phi}) \mathbf{R}_{T}^{\frac{1}{2}}\mathbf{F}\mathbf{F}^\HH\mathbf{R}_{T}^{\frac{1}{2}} \right)^{-1}.$
			\begin{equation}
				\begin{split}
					\mathbf{M}_{T,j}=\left(\mathbf{I}_{N_S}+\alpha_j \mathbf{R}_{T}^{\frac{1}{2}}\mathbf{\Phi}\mathbf{R}_{T}^{\frac{1}{2}} \right)^{-1}.
				\end{split}
			\end{equation}
			According to \eqref{fixeq1}, we have
			\begin{equation}
				\begin{split}
					-\alpha_j^2\tr\left(\mathbf{M}_{T,j}\mathbf{T}\right)+N_S \alpha_j - \frac{N_S \lambda_{R,j}}{\left(1+\lambda_{R,j}\delta(\lambda_{R,j})\right)^2} = 0.
				\end{split}
			\end{equation}
			It indicates that $	\nabla_{\mathbf{\Phi}}\bar{\varrho}_{j}(\mathbf{\Phi})= \alpha_j \mathbf{R}_T^{\frac{1}{2}}\mathbf{M}_{T,j}\mathbf{R}_T^{\frac{1}{2}}$. 
%				\begin{align}
%						\nabla_{\mathbf{\Phi}}\bar{\varrho}_{j}(\mathbf{\Phi})= \alpha_j \mathbf{R}_T^{\frac{1}{2}}\mathbf{M}_{T,j}\mathbf{R}_T^{\frac{1}{2}}.
%					\end{align}

			%	
			\section{Proof of Proposition \ref{PropEQ2}}
			\label{proofPropEQ2}
			The $(m,n)$th entry of $\nabla_{\mathbf{\Phi}}\bar{\varkappa}_{j}(\mathbf{\Phi})$ is given by
			\begin{equation}
				\begin{split}\nonumber
					&\left[\nabla_{\mathbf{\Phi}}\bar{\varkappa}_{j}(\mathbf{\Phi})\right]_{m,n}\triangleq \frac{\partial \sigma_s^2\lambda_{R,j} \tr \left(\mathbf{R}_T \left(\mathbf{I}_{N_T}+\alpha_j \mathbf{T}(\mathbf{\Phi})\right)^{-1}\right)}{\partial \Omega_{m,n}^*}\\
					&=-\sigma_s^2\lambda_{R,j} \tr  \left(\mathbf{R}_T \mathbf{M}_{T,j}
					\left(\alpha_{m,n}'(\lambda_{R,j}) \mathbf{T} + \alpha_j [{\mathbf{T}'}]_{m,n} \right)\mathbf{M}_{T,j}\right)\\
					&\overset{(a)}{=}\sigma_s^2\lambda_{R,j} \alpha_j^2\tr\left(\mathbf{R}_T \mathbf{M}_{T,j}\mathbf{T} \mathbf{M}_{T,j}\right) [ \mathbf{\Delta}_{\lambda_{R,j}}'(\mathbf{\Phi})]_{m,n}\\
					&\quad-\sigma_s^2\lambda_{R,j}\alpha_j[\mathbf{R}_T^{\frac{1}{2}}\mathbf{M}_{T,j}\mathbf{R}_T\mathbf{M}_{T,j}\mathbf{R}_T^{\frac{1}{2}}]_{m,n},\\
				\end{split}
			\end{equation}
			where  (a) follows from \eqref{derproof001} and \eqref{derproof002}.
			Thus, the gradient of $\bar{\varkappa}_{j}(\mathbf{\Phi})$ with respect to $\mathbf{F}$ is given by
			\begin{equation}
				\begin{split}
					&\nabla_{\mathbf{\Phi}}\bar{\varkappa}_{j}(\mathbf{\Phi})\triangleq \frac{\partial \bar{\varkappa}_{j}(\mathbf{\Phi})}{\partial \mathbf{\Omega}^*}\\
					&=\sigma_s^2\lambda_{R,j} \alpha_j^2\tr\left(\mathbf{R}_T \mathbf{M}_{T,j}\mathbf{T} \mathbf{M}_{T,j}\right)  \mathbf{\Delta}_{\lambda_{R,j}}'(\mathbf{\Phi})\\
					&\quad-\sigma_s^2\lambda_{R,j}\alpha_j\mathbf{R}_T^{\frac{1}{2}}\mathbf{M}_{T,j}\mathbf{R}_T\mathbf{M}_{T,j}\mathbf{R}_T^{\frac{1}{2}}.\\
					%				&=\sigma_s^2\lambda_{R,j} \tr\left(\nabla_{\mathbf{\Phi}}\bar{\varrho}_{j}(\mathbf{\Phi})\nabla_{\mathbf{\Phi}}^\HH\bar{\varrho}_{j}(\mathbf{\Phi})\right)  \mathbf{\Delta}_{\lambda_{R,j}}'(\mathbf{\Phi})\\
					%				&\quad-\sigma_s^2\lambda_{R,j}\mathbf{R}_T^{\frac{1}{2}}\mathbf{M}_{T,j}\mathbf{R}_T^{\frac{1}{2}}\nabla_{\mathbf{\Phi}}\bar{\varrho}_{j}(\mathbf{\Phi}). 
				\end{split}
			\end{equation}


\begin{thebibliography}{10}
\providecommand{\url}[1]{#1}
\csname url@samestyle\endcsname
\providecommand{\newblock}{\relax}
\providecommand{\bibinfo}[2]{#2}
\providecommand{\BIBentrySTDinterwordspacing}{\spaceskip=0pt\relax}
\providecommand{\BIBentryALTinterwordstretchfactor}{4}
\providecommand{\BIBentryALTinterwordspacing}{\spaceskip=\fontdimen2\font plus
\BIBentryALTinterwordstretchfactor\fontdimen3\font minus
  \fontdimen4\font\relax}
\providecommand{\BIBforeignlanguage}[2]{{%
\expandafter\ifx\csname l@#1\endcsname\relax
\typeout{** WARNING: IEEEtran.bst: No hyphenation pattern has been}%
\typeout{** loaded for the language `#1'. Using the pattern for}%
\typeout{** the default language instead.}%
\else
\language=\csname l@#1\endcsname
\fi
#2}}
\providecommand{\BIBdecl}{\relax}
\BIBdecl

\bibitem{xie2023sensing}
L.~Xie, F.~Liu, Z.~Xie, Z.~Jiang, and S.~Song, ``Sensing mutual information
  with random signals in {G}aussian channels,'' \emph{arXiv preprint
  arXiv:2311.07081}, 2023.

\bibitem{liu2020joint}
F.~Liu, C.~Masouros, A.~P. Petropulu, H.~Griffiths, and L.~Hanzo, ``Joint radar
  and communication design: Applications, state-of-the-art, and the road
  ahead,'' \emph{IEEE Trans. Commun.}, vol.~68, no.~6, pp. 3834--3862, 2020.

\bibitem{liu2022integrated}
F.~Liu, Y.~Cui, C.~Masouros, J.~Xu, T.~X. Han, Y.~C. Eldar, and S.~Buzzi,
  ``Integrated sensing and communications: Toward dual-functional wireless
  networks for {6G} and beyond,'' \emph{IEEE J. Sel. Areas Commun.}, vol.~40,
  no.~6, pp. 1728--1767, 2022.

\bibitem{liu2022survey}
A.~Liu, Z.~Huang, M.~Li, Y.~Wan, W.~Li, T.~X. Han, C.~Liu, R.~Du, D.~K.~P. Tan,
  J.~Lu \emph{et~al.}, ``A survey on fundamental limits of integrated sensing
  and communication,'' \emph{IEEE Commun. Surv. Tutor.}, vol.~24, no.~2, pp.
  994--1034, 2022.

\bibitem{xie2023collaborative}
L.~Xie, S.~Song, Y.~C. Eldar, and K.~B. Letaief, ``Collaborative sensing in
  perceptive mobile networks: Opportunities and challenges,'' \emph{IEEE Wirel.
  Commun.}, vol.~30, no.~1, pp. 16--23, 2023.

\bibitem{xie2023networked}
L.~Xie, S.~Song, and K.~B. Letaief, ``Networked sensing with ai-empowered
  interference management: Exploiting macro-diversity and array gain in
  perceptive mobile networks,'' \emph{IEEE J. Sel. Areas Commun.}, vol.~41,
  no.~12, pp. 3863--3877, 2023.

\bibitem{liu2021cramer}
F.~Liu, Y.-F. Liu, A.~Li, C.~Masouros, and Y.~C. Eldar, ``Cram{\'e}r-{R}ao
  bound optimization for joint radar-communication beamforming,'' \emph{IEEE
  Trans. Signal Process.}, vol.~70, pp. 240--253, 2021.

\bibitem{xie2022perceptive}
L.~Xie, P.~Wang, S.~Song, and K.~B. Letaief, ``Perceptive mobile network with
  distributed target monitoring terminals: Leaking communication energy for
  sensing,'' \emph{IEEE Trans. Wirel. Commun.}, vol.~21, no.~12, pp.
  10\,193--10\,207, 2022.

\bibitem{herbert2017mmse}
S.~Herbert, J.~R. Hopgood, and B.~Mulgrew, ``{MMSE} adaptive waveform design
  for active sensing with applications to {MIMO} radar,'' \emph{IEEE Trans.
  Signal Process.}, vol.~66, no.~5, pp. 1361--1373, 2017.

\bibitem{hachem2008new}
W.~Hachem, O.~Khorunzhiy, P.~Loubaton, J.~Najim, and L.~Pastur, ``A new
  approach for mutual information analysis of large dimensional multi-antenna
  channels,'' \emph{IEEE Trans. Inf. Theory}, vol.~54, no.~9, pp. 3987--4004,
  2008.

\bibitem{jeruchim1984techniques}
M.~Jeruchim, ``Techniques for estimating the bit error rate in the simulation
  of digital communication systems,'' \emph{IEEE J. Sel. Areas Commun.},
  vol.~2, no.~1, pp. 153--170, 1984.

\bibitem{ko2000outage}
Y.-C. Ko, M.-S. Alouini, and M.~K. Simon, ``Outage probability of diversity
  systems over generalized fading channels,'' \emph{IEEE Trans. commun.},
  vol.~48, no.~11, pp. 1783--1787, 2000.

\bibitem{bell1993information}
M.~R. Bell, ``Information theory and radar waveform design,'' \emph{IEEE Trans.
  Inf. Theory}, vol.~39, no.~5, pp. 1578--1597, 1993.

\bibitem{tang2018spectrally}
B.~Tang and J.~Li, ``Spectrally constrained {MIMO} radar waveform design based
  on mutual information,'' \emph{IEEE Trans. Signal Process.}, vol.~67, no.~3,
  pp. 821--834, 2018.

\bibitem{yang2007mimo}
Y.~Yang and R.~S. Blum, ``{MIMO} radar waveform design based on mutual
  information and minimum mean-square error estimation,'' \emph{IEEE Trans.
  Aerosp. electro. syst.}, vol.~43, no.~1, pp. 330--343, 2007.

\bibitem{li200824}
Z.~Li and K.~Wu, ``24-{GHz} frequency-modulation continuous-wave radar
  front-end system-on-substrate,'' \emph{IEEE Trans. Microw. Theory Tech.},
  vol.~56, no.~2, pp. 278--285, 2008.

\bibitem{wang2014application}
G.~Wang, J.-M. Munoz-Ferreras, C.~Gu, C.~Li, and R.~Gomez-Garcia, ``Application
  of linear-frequency-modulated continuous-wave ({LFMCW}) radars for tracking
  of vital signs,'' \emph{IEEE Trans. Microw. Theory Tech.}, vol.~62, no.~6,
  pp. 1387--1399, 2014.

\bibitem{huang2019long}
P.~Huang, X.-G. Xia, G.~Liao, Z.~Yang, and Y.~Zhang, ``Long-time coherent
  integration algorithm for radar maneuvering weak target with acceleration
  rate,'' \emph{IEEE Trans. Geosci. Remote Sens.}, vol.~57, no.~6, pp.
  3528--3542, 2019.

\bibitem{xie2020recursive}
L.~Xie, Z.~He, J.~Tong, and W.~Zhang, ``A recursive angle-doppler channel
  selection method for reduced-dimension space-time adaptive processing,''
  \emph{IEEE Trans. Aerosp. Electron. Syst.}, vol.~56, no.~5, pp. 3985--4000,
  2020.

\bibitem{liu2023deterministic}
F.~Liu, Y.~Xiong, K.~Wan, T.~X. Han, and G.~Caire, ``Deterministic-random
  tradeoff of integrated sensing and communications in gaussian channels: A
  rate-distortion perspective,'' in \emph{2023 IEEE International Symposium on
  Information Theory (ISIT)}.\hskip 1em plus 0.5em minus 0.4em\relax IEEE,
  2023, pp. 2326--2331.

\bibitem{dong2023rethinking}
F.~Dong, F.~Liu, S.~Lu, and Y.~Xiong, ``Rethinking estimation rate for wireless
  sensing: A rate-distortion perspective,'' \emph{IEEE Trans. Veh. Technol.},
  vol.~72, no.~12, pp. 16\,876--16\,881, 2023.

\bibitem{xiong2023fundamental}
Y.~Xiong, F.~Liu, Y.~Cui, W.~Yuan, T.~X. Han, and G.~Caire, ``On the
  fundamental tradeoff of integrated sensing and communications under
  {G}aussian channels,'' \emph{IEEE Trans. Inf. Theory}, 2023.

\bibitem{xiong2023generalized}
Y.~Xiong, F.~Liu, and M.~Lops, ``Generalized deterministic-random tradeoff in
  integrated sensing and communications: The sensing-optimal operating point,''
  \emph{IEEE International Conference on Acoustics, Speech and Signal
  Processing}, Accepted, 2024.

\bibitem{lu2023sensing}
S.~Lu, F.~Liu, F.~Dong, Y.~Xiong, J.~Xu, and Y.-F. Liu, ``Sensing with random
  signals,'' \emph{IEEE International Conference on Acoustics, Speech and
  Signal Processing}, Accepted, 2024.

\bibitem{lu2023random}
S.~Lu, F.~Liu, F.~Dong, Y.~Xiong, J.~Xu, Y.-F. Liu, and S.~Jin, ``Random {ISAC}
  signals deserve dedicated precoding,'' \emph{arXiv preprint
  arXiv:2311.01822}, 2023.

\bibitem{1021913}
J.~Kermoal, L.~Schumacher, K.~Pedersen, P.~Mogensen, and F.~Frederiksen, ``A
  stochastic {MIMO} radio channel model with experimental validation,''
  \emph{IEEE J. Sel. Areas Commun.}, vol.~20, no.~6, pp. 1211--1226, 2002.

\bibitem{1300860}
K.~Yu, M.~Bengtsson, B.~Ottersten, D.~McNamara, P.~Karlsson, and M.~Beach,
  ``Modeling of wide-band {MIMO} radio channels based on {NLoS} indoor
  measurements,'' \emph{IEEE Trans. Veh. Technol.}, vol.~53, no.~3, pp.
  655--665, 2004.

\bibitem{5432999}
Y.~I. Abramovich, G.~J. Frazer, and B.~A. Johnson, ``Iterative adaptive
  kronecker mimo radar beamformer: Description and convergence analysis,''
  \emph{IEEE Trans. Signal Process.}, vol.~58, no.~7, pp. 3681--3691, 2010.

\bibitem{ren2023fundamental}
Z.~Ren, Y.~Peng, X.~Song, Y.~Fang, L.~Qiu, L.~Liu, D.~W.~K. Ng, and J.~Xu,
  ``Fundamental {CRB}-rate tradeoff in multi-antenna {ISAC} systems with
  information multicasting and multi-target sensing,'' \emph{IEEE Trans. Wirel.
  Commun.}, 2023.

\bibitem{325008}
J.~Helferty and D.~Mudgett, ``Optimal observer trajectories for bearings only
  tracking by minimizing the trace of the {C}ramer-{R}ao lower bound,'' in
  \emph{Proc. 32th IEEE Conf. Decis. Control}, 1993, pp. 936--939 vol.1.

\bibitem{yan2015simultaneous}
J.~Yan, H.~Liu, B.~Jiu, B.~Chen, Z.~Liu, and Z.~Bao, ``Simultaneous multibeam
  resource allocation scheme for multiple target tracking,'' \emph{IEEE Trans.
  Signal Process.}, vol.~63, no.~12, pp. 3110--3122, 2015.

\bibitem{shi2011iteratively}
Q.~Shi, M.~Razaviyayn, Z.-Q. Luo, and C.~He, ``An iteratively weighted mmse
  approach to distributed sum-utility maximization for a {MIMO} interfering
  broadcast channel,'' \emph{IEEE Trans. Signal Process.}, vol.~59, no.~9, pp.
  4331--4340, 2011.

\bibitem{reeves2018mutual}
G.~Reeves, H.~D. Pfister, and A.~Dytso, ``Mutual information as a function of
  matrix {SNR} for linear gaussian channels,'' in \emph{2018 IEEE International
  Symposium on Information Theory (ISIT)}.\hskip 1em plus 0.5em minus
  0.4em\relax IEEE, 2018, pp. 1754--1758.

\bibitem{absil2009optimization}
P.-A. Absil, R.~Mahony, and R.~Sepulchre, \emph{Optimization algorithms on
  matrix manifolds}.\hskip 1em plus 0.5em minus 0.4em\relax Princeton
  University Press, 2008.

\bibitem{boyd2004convex}
S.~P. Boyd and L.~Vandenberghe, \emph{Convex optimization}.\hskip 1em plus
  0.5em minus 0.4em\relax Cambridge university press, 2004.

\bibitem{grant2014cvx}
M.~Grant and S.~Boyd, ``{CVX}: Matlab software for disciplined convex
  programming, version 2.1,'' 2014.

\bibitem{bibby1974axiomatisations}
J.~Bibby, ``Axiomatisations of the average and a further generalisation of
  monotonic sequences,'' \emph{Glasgow Mathematical Journal}, vol.~15, no.~1,
  pp. 63--65, 1974.

\bibitem{6834753}
M.~R. Akdeniz, Y.~Liu, M.~K. Samimi, S.~Sun, S.~Rangan, T.~S. Rappaport, and
  E.~Erkip, ``Millimeter wave channel modeling and cellular capacity
  evaluation,'' \emph{IEEE J. Sel. Areas Commun.}, vol.~32, no.~6, pp.
  1164--1179, 2014.

\end{thebibliography}
		\end{document}